\documentclass{aa}
\usepackage{psfig,txfonts}

\def\ltsima{$\; \buildrel < \over \sim \;$}
\def\lsim{\lower.5ex\hbox{\ltsima}}
\def\gtsima{$\; \buildrel > \over \sim \;$}
\def\gsim{\lower.5ex\hbox{\gtsima}}
\def\fdeg{\hbox{$\,.\!\!^{\circ}$}}

\begin{document}

\title{Optical and near-infrared observations \\ of the GRB020405 
afterglow\thanks{Based on observations made with ESO telescopes at 
Paranal and La Silla Observatories under programme ID 69.D-0701, 
with the telescopes TNG, WHT and JKT, operating in the Spanish
Observatorio del Roque de los Muchachos of the Instituto de
Astrof\'{\i}sica de Canarias, and with the 1m telescope of SO in 
NainiTal, India}}

\author{N. Masetti\inst{1} 
\and
E. Palazzi\inst{1}
\and
E. Pian\inst{1,2}
\and
A. Simoncelli\inst{3}
\and
L.K. Hunt\inst{4}
\and
E. Maiorano\inst{1,3}
\and
A. Levan\inst{5}
\and
L. Christensen\inst{6}
\and
E. Rol\inst{7}
\and
S. Savaglio\inst{8,9}
\and
R. Falomo\inst{10}
\and
A.J. Castro-Tirado\inst{11}
\and
J. Hjorth\inst{12}
\and
A. Delsanti\inst{13}
\and
M. Pannella\inst{14}
\and
V. Mohan\inst{15}
\and
S.B. Pandey\inst{15}
\and
R. Sagar\inst{15}
\and
L. Amati\inst{1}
\and
I. Burud\inst{16}
\and
J.M. Castro Cer\'on\inst{17}
\and
F. Frontera\inst{1,18}
\and
A.S. Fruchter\inst{16}
\and
J.P.U. Fynbo\inst{19}
\and
J. Gorosabel\inst{11}
\and
L. Kaper\inst{7}
\and
S. Klose\inst{20}
\and
C. Kouveliotou\inst{21}
\and
L. Nicastro\inst{22}
\and
H. Pedersen\inst{12}
\and
J. Rhoads\inst{16}
\and
I. Salamanca\inst{7}
\and
N. Tanvir\inst{23}
\and
P.M. Vreeswijk\inst{7,24}
\and
R.A.M.J. Wijers\inst{7}
\and
E.P.J. van den Heuvel\inst{7}
}

\institute{
Istituto di Astrofisica Spaziale e Fisica Cosmica -- Sezione di Bologna, 
CNR, via Gobetti 101, I-40129 Bologna, Italy
\and
INAF -- Osservatorio Astronomico di Trieste, via G.B. Tiepolo 11, I-34131
Trieste, Italy
\and
Dipartimento di Astronomia, Universit\`a di Bologna, via Ranzani 1,
I-40126 Bologna, Italy
\and
Istituto di Radioastronomia -- Sezione di Firenze, CNR, largo E. Fermi 5,
I-50125, Florence, Italy
\and
Department of Physics and Astronomy, University of Leicester, University
Road, Leicester, LE1 7RH, United Kingdom
\and
Astrophysikalisches Institut, 14482 Potsdam, Germany
\and
Institute of Astronomy ``Anton Pannekoek", University of Amsterdam,
Kruislaan 403, 1098 SJ Amsterdam, The Netherlands
\and
The Johns Hopkins University, 3400 North Charles Street, Bartimore, MD
21218, USA
\and
INAF -- Osservatorio Astronomico di Roma, via Frascati 33,
I-00040 Monteporzio Catone, Italy
\and
INAF -- Osservatorio Astronomico di Padova, vicolo dell'Osservatorio 5,
I-35122 Padua, Italy
\and
Instituto de Astrof\'{\i}sica de Andaluc\'{\i}a (IAA-CSIC),
P.O. Box 03004, E-18080 Granada, Spain
\and
Astronomical Observatory, University of Copenhagen, Juliane Maries Vej 30,
DK--2100 Copenhagen \O, Denmark
\and
Observatoire de Paris-Meudon - LESIA, 5 Place Jules Janssen, 92195 
Meudon, France
\and
Max-Planck-Institut f\"ur Extraterrestrische Physik, 85741 Garching,
Germany
\and
State Observatory, Manora Peak, NainiTal -- 263129 Uttaranchal, India
\and
Space Telescope Science Institute, 3700 San Martin Drive, Baltimore, MD
21218, USA
\and
Real Instituto y Observatorio de la Armada, Secci\'on de Astronom\'\i a,
11.110 San Fernando-Naval (C\'adiz), Spain
\and
Dipartimento di Fisica, Universit\`a di Ferrara, via Paradiso 12, I-44100
Ferrara, Italy
\and
Department of Physics and Astronomy, University of \AA rhus, Ny
Munkegade, 8000 \AA rhus C, Denmark
\and
Th\"uringer Landessternwarte Tautenburg, D-07778 Tautenburg, Germany
\and
NASA MSFC, SD-50, Huntsville, AL 35812, USA
\and
Istituto di Astrofisica Spaziale e Fisica Cosmica -- Sezione di
Palermo, CNR, via La Malfa 153, I-90146 Palermo, Italy
\and
Department of Physical Sciences, University of Hertfordshire,
College Lane, Hatfield, Herts AL10 9AB, UK
\and
European Southern Observatory, Casilla 19001, Santiago 19, Chile
}

\titlerunning{Optical-NIR monitoring of the GRB020405 afterglow}
\authorrunning{Masetti et al.}

\offprints{N. Masetti, {\tt masetti@bo.iasf.cnr.it}}

\date{Received 19 February 2003; Accepted 27 March 2003}

\abstract{We report on photometric, spectroscopic and polarimetric 
monitoring of the optical and near-infrared (NIR) afterglow of GRB020405.
Ground-based optical observations, performed with 8 different telescopes,
started about 1 day after the high-energy prompt
event and spanned a period of $\sim$10 days; the addition of archival HST
data extended the coverage up to $\sim$150 days after the GRB. We report
the first detection of the afterglow in NIR bands.
The detection of Balmer and oxygen emission lines in the optical spectrum
of the host galaxy indicates that the GRB is located at redshift $z$ =
0.691. Fe {\sc ii} and Mg {\sc ii} absorption systems are detected at $z$
= 0.691 and at $z$ = 0.472 in the afterglow optical spectrum. The latter
system is likely caused by absorbing clouds in the galaxy complex located
$\sim$2$''$ southwest of the GRB020405 host. Hence, for the first time,
the galaxy responsible for an intervening absorption line system in the
spectrum of a GRB afterglow is spectroscopically identified.
Optical and NIR photometry of the afterglow indicates that, between 1 and
10 days after the GRB, the decay in all bands is consistent with a 
single power law of index $\alpha$ = 1.54$\pm$0.06. The late-epoch
VLT $J$-band and HST optical points lie above the extrapolation of this
power law, so that a plateau (or ``bump") is apparent in the $VRIJ$ light
curves at 10-20 days after the GRB. The light curves at epochs later than
day $\sim$20 after the GRB are consistent with a power-law decay with
index $\alpha'$ = 1.85$\pm$0.15. While other authors have proposed to
reproduce the bump with the template of the supernova (SN) 1998bw, 
considered the prototypical `hypernova', we suggest that it can also be 
modeled with a SN having the same temporal profile as the other 
proposed hypernova SN2002ap, but 1.3 mag brighter at peak, and located at 
the GRB redshift.
Alternatively, a shock re-energization may be responsible for the
rebrightening. A single polarimetric $R$-band measurement shows that the
afterglow is polarized, with $P$ = 1.5$\pm$0.4 \% and polarization angle
$\theta$ = 172$^{\circ}$$\pm$8$^{\circ}$. Broad-band optical-NIR spectral
flux distributions show, in the first days after the GRB, a change of
slope across the $J$ band which we interpret as due to the presence of the
electron cooling frequency $\nu_{\rm c}$. The analysis of the
multiwavelength spectrum within the standard fireball model suggests that
a population of relativistic electrons with index $p \sim$ 2.7 produces
the optical-NIR emission via synchrotron radiation in an adiabatically
expanding blastwave, with negligible host galaxy extinction, and the
X--rays via Inverse Compton scattering off lower-frequency afterglow
photons.

}

\maketitle

\section{Introduction}

GRB020405 was detected by the Third InterPlanetary Network (IPN) on 2002 
April 5.02877 UT with a duration of $\sim$40 s and localized to an error
box of 75 square arcmin size. In the 25--100 keV band it had a total
fluence of $\sim$3 $\times 10^{-5}$ erg cm$^{-2}$ and a peak flux of $\sim
10^{-6}$ erg cm$^{-2}$ s$^{-1}$ (Hurley et al. 2002).  This GRB was also
observed by the GRBM onboard {\it BeppoSAX}, with a duration of $\sim$60 s
in the 40--700 keV band and a 50--700 keV fluence of
$\sim$4$\times$10$^{-5}$ erg cm$^{-2}$ (Price et al. 2003).

Approximately 18 hours after the GRB, Price et al. (2002a, 2003) detected
a relatively bright ($R \sim$ 18.5) source within the IPN error box which
was not present on the DSS-II red plate. This Optical Transient (OT),
located at coordinates (J2000) RA = 13$^{\rm h}$ 58$^{\rm m}$ 03$\fs$12;
Dec = $-$31$^{\circ}$ 22$'$ 22$\farcs$2 (with an error of 0$\farcs$3 along
both directions), was confirmed by subsequent observations and identified
as the afterglow of GRB020405 (Castro-Tirado et al. 2002; Palazzi et al.
2002; Hjorth et al. 2002; Price et al. 2002b; Gal-Yam et al. 2002a; Covino
et al. 2002a,b). Optical spectroscopy allowed Masetti et al. (2002a) and
Price et al. (2003) to determine the redshift of the GRB, $z$ = 0.691
(see also Sect. 3.3). 

A counterpart to the GRB has been detected also at radio and X--ray 
wavelengths (Mirabal et al. 2002; Berger et al. 2003).
By fitting to the optical light curves a power law $F$($t$) $\propto
t^{-\alpha}$, typical of GRB afterglows, Price et al. (2002b) and Covino
et al. (2002b) found $\alpha \sim$ 1.26 and $\alpha$ = 1.52$\pm$0.12,
respectively, from observations performed within 3 days after the GRB. 
By using a data set spanning $\sim$5 days, Price et al. 
(2003) found instead that a better fit to the early optical afterglow
light curve was obtained with a smoothly broken power law with decay
indices $\alpha_1 \sim$ 0.9 and $\alpha_2 \sim$ 1.9 before and after a
break which occurred $\sim$1.7 days after the GRB, respectively.
On the other hand, Bersier et al. (2003) found no evidence of a break
between 1.24 and 4.3 days after the GRB in their $R$-band data, which
followed a single power-law decay with index $\alpha \sim$ 1.7. 

Polarimetric measurements by Covino et al. (2003) and Bersier et al.
(2003) suggest early rapid variability of the percentage of linear
polarization between values of $\sim$1.5\% and $\sim$10\%.

Price et al. (2003), using late-time HST observations covering the time
interval between 20 and 140 days after the high-energy prompt event, found
a ``red bump'', or flattening, in the optical light curves of the OT
and suggested that this could be due to an emerging supernova (SN)
component, which they modeled with the template of SN1998bw located at the
redshift of GRB020405 and further dimmed by about 0.5 mag. A similar
assumption was made by Dado et al. (2002) in their modeling of the
GRB020405 optical afterglow within the cannonball picture.

Optical observations also revealed the presence of a relatively large
(about 2$''$$\times$1$''$, elongated in the N-S direction) nebulosity
located $\sim$2$''$ southwest of the OT and interpreted as its putative
host galaxy (Hjorth et al. 2002). 

In this paper we report on optical imaging, spectroscopy and polarimetry,
along with the first detection in the near-infrared (NIR) bands of the
GRB020405 afterglow. The data were acquired, in the framework of the 
GRACE\footnote{{\sl GRB Afterglow Collaboration at ESO}: see the web
page \\ {\tt http://zon.wins.uva.nl/grb/\~{}grace}}
collaboration, at the ESO telescopes of Cerro Paranal
and La Silla (Chile) starting on 6 April 2002, i.e., $\sim$1 day after 
the high-energy prompt event. Observations from
the 1-metre SO telescope, WHT and TNG are also included, along with
the reanalysis of the public HST observations of this GRB.

The paper is organized as follows: Sect. 2 describes the
optical and NIR observations and the data analysis; the results are reported
in Sect. 3 and discussed in Sect. 4; in Sect. 5 we report our 
conclusions. Throughout the paper we assume a cosmology with 
$H_{\rm 0}$ = 65 km s$^{-1}$ Mpc$^{-1}$, $\Omega_{\Lambda}$ = 0.7 and
$\Omega_{\rm m}$ = 0.3; also, when not otherwise stated, 
uncertainties will be reported at 1$\sigma$ confidence level, and upper 
limits at 3$\sigma$ confidence level.

\section{Observations and data reduction}

\subsection{Optical and NIR photometry}

Optical $BVRI$ observations in the Johnson-Cousins photometric system were
performed with VLT-{\it Melipal} plus FORS1, VLT-{\it Yepun} plus FORS2,
NTT plus SUSI2 and EMMI, and 1.54-metre Danish telescope plus DFOSC. 

FORS1 is equipped with a 2048$\times$2048 pixels Tektronix CCD 
which covers a 6$\farcm$8$\times$6$\farcm$8 field in the standard
resolution imaging mode with a scale of 0$\farcs$2 pix$^{-1}$;
FORS2 is equipped with a mosaic of two 2048$\times$4096 
pixels MIT CCDs, usually operating in 2$\times$2 standard binning mode, 
thus covering a field with size 6$\farcm$8$\times$6$\farcm$8 with a 
spatial resolution of 0$\farcs$25 pix$^{-1}$ in the standard resolution 
imaging mode.
SUSI2 incorporates a mosaic of two 2048$\times$4096 pixels EEV CCDs
which cover a field of 5$\farcm$5$\times$5$\farcm$5, corresponding to
0$\farcs$08 pix$^{-1}$; EMMI (Red Arm) carried until 14 May 2002 a
2048$\times$2047 pixels Tektronix CCD with a scale of 0$\farcs$27
pix$^{-1}$ and a field coverage of 9$\farcm$15$\times$8$\farcm$6 in size.
DFOSC is equipped with a 2048$\times$2048 pixels CCD covering a 
13$\farcm$7$\times$13$\farcm$7 field, thus securing a spatial
resolution of 0$\farcs$39 pix$^{-1}$.

Optical observations were also obtained on 5 and 6 April 2002
($I$ band) at the 1.0m Sampurnanand telescope of SO located in 
Nainital (India), and on 6 April 2002 at the Canary Islands (Spain) 
with the 4.2m WHT plus PFIP ($UBVRI$ bands) and with the 3.58m TNG 
plus DOLoReS ($V$ band).
The 1.0m SO telescope carries a 2048$\times$2048 pixels CCD, with a 
13$'$$\times$13$'$ field of view; in its standard 2$\times$2 binning 
mode it has a spatial resolution of 0$\farcs$76 pix$^{-1}$.
The imaging camera PFIP carries two 2100$\times$4200 EEV CCDs
which cover a field of view of 16$\farcm$2$\times$16$\farcm$2,
giving a plate scale of 0$\farcs$24 pix$^{-1}$; the spectrophotometer 
DOLoReS carries a 2048$\times$2048 pixels Loral CCD which can image a 
field of 9$\farcm$5$\times$9$\farcm$5 with a scale of 0$\farcs$275 
pix$^{-1}$.

NIR imaging was obtained at ESO NTT with SofI ($J$, $H$
and $K_s$ bands) and at VLT-{\it Antu} with ISAAC ($J_s$ band).
The SofI infrared spectrograph and imaging camera works in the
0.9--2.5 $\mu$m NIR range by using a Hawaii 1024$\times$1024 pixel
HgCdTe array. In the small-field imaging mode the plate scale is
0$\farcs$144 pix$^{-1}$ and the corresponding field of view is
2$\farcm$4$\times$2$\farcm$4. ISAAC is equipped, in the 0.9--2.5 $\mu$m
range, with a Rockwell Hawaii 1024$\times$1024 pixel HgCdTe array which
has a scale of 0$\farcs$148 pix$^{-1}$ and secures imaging with a field of
view of 2$\farcm$5$\times$2$\farcm$5.
The $J_s$ filter is centered at 1.24 $\mu$m and has a full width at half
maximum of 0.16 $\mu$m; its overall response is 25\% lower than the $J$ 
filter. This is confirmed by comparing the nightly zero-point 
coefficients of $J$ and $J_s$ frames in our data set. We thus reported 
the ISAAC $J_s$ magnitudes to the standard $J$ band by increasing the 
observed $J_s$-band flux densities by 25\%.
In order to allow for sky subtraction, the total integration time of
each NIR pointing was split into dithered images of 15 s each when using
SofI, and of 30 s each during ISAAC observations. In both cases the
dithering was 40$''$ between consecutive images.

The complete log of our optical and NIR imaging observations 
is reported in Table 1.

\begin{table*}[t!]
\caption[]{Journal of the GRB020405 afterglow ground-based optical and NIR
photometric observations. Magnitudes are not corrected for Galactic 
interstellar absorption}
\begin{center}
\begin{tabular}{lcccccl}
\noalign{\smallskip}
\hline
\noalign{\smallskip}
\multicolumn{1}{c}{Mid-exposure} & Telescope & Instrument & Filter & Total 
exposure & Seeing & \multicolumn{1}{c}{Magnitude}\\
\multicolumn{1}{c}{time (UT)} & & & & time (s) & (arcsec) &  \\
\noalign{\smallskip}
\hline
\noalign{\smallskip}
2002 Apr 5.812  & 1m SO      & CCD     & $I$ & 2$\times$600 & 2.6 & 
	19.30$\pm$0.06\\
2002 Apr 6.069  & TNG          & DOLoRes & $V$ & 2$\times$300 & 1.8 & 
	20.66$\pm$0.02$^*$\\
2002 Apr 6.013  & WHT          & PFIP    & $R$ & 300 & 2.3 & 20.17$\pm$0.03\\
2002 Apr 6.109  & WHT          & PFIP    & $U$ & 900 & 2.2 & 20.48$\pm$0.06\\
2002 Apr 6.118  & WHT          & PFIP    & $B$ & 600 & 2.5 & 21.21$\pm$0.03\\
2002 Apr 6.124  & WHT          & PFIP    & $V$ & 300 & 2.0 & 20.73$\pm$0.03\\
2002 Apr 6.129  & WHT          & PFIP    & $R$ & 300 & 2.8 & 20.28$\pm$0.04\\
2002 Apr 6.134  & WHT          & PFIP    & $I$ & 300 & 3.2 & 19.61$\pm$0.08\\
2002 Apr 6.136  & 1.54D        & DFOSC   & $R$ & 600 & 1.4 & 20.38$\pm$0.02\\
2002 Apr 6.186  & {\it Melipal}& FORS1   & $R$ &  60 & 0.7 & 20.48$\pm$0.01\\
2002 Apr 6.197  & {\it Melipal}& FORS1   & $R$ &  30 & 0.8 & 20.46$\pm$0.01\\
2002 Apr 6.211  & 1.54D        & DFOSC   & $R$ & 600 & 1.4 & 20.44$\pm$0.02\\
2002 Apr 6.220  & 1.54D        & DFOSC   & $V$ & 600 & 1.4 & 20.81$\pm$0.02\\
2002 Apr 6.220  & NTT          & SofI   & $H$ & 300 & 1.0 & 18.07$\pm$0.07\\
2002 Apr 6.224  & NTT          & SofI   & $J$ & 300 & 1.0 & 18.69$\pm$0.06\\
2002 Apr 6.230  & 1.54D        & DFOSC   & $B$ & 600 & 1.4 & 21.31$\pm$0.03\\
2002 Apr 6.230  & NTT          & SofI   & $K_s$ &  600 & 0.9 &
	17.34$\pm$0.06\\
2002 Apr 6.234  & {\it Melipal}& FORS1   & $R$ &  20 & 0.8 & 20.53$\pm$0.02\\
2002 Apr 6.246  & 1.54D        & DFOSC   & $Gunn~i$ & 600 & 1.1 &
	19.88$\pm$0.02\\
2002 Apr 6.797  & 1m SO      & CCD     & $I$ & 5$\times$600 & 2.5 & $>$20.0\\
2002 Apr 7.132  & 1.54D        & DFOSC   & $R$ & 600 & 1.2 & 21.20$\pm$0.04\\
2002 Apr 7.240  & 1.54D        & DFOSC   & $R$ & 600 & 0.9 & 21.30$\pm$0.06\\
2002 Apr 7.249  & 1.54D        & DFOSC   & $V$ & 600 & 1.1 & 21.79$\pm$0.05\\
2002 Apr 7.266  & 1.54D        & DFOSC   & $B$ & 600 & 1.2 & 22.29$\pm$0.07\\
2002 Apr 7.270  & NTT          & SofI   & $H$ & 600 & 0.8 & 18.90$\pm$0.07\\
2002 Apr 7.278  & NTT          & SofI   & $J$ & 600 & 0.8 & 19.55$\pm$0.06\\
2002 Apr 7.281  & {\it Melipal}& FORS1   & $V$ & 2$\times$30 & 0.7 &
	21.95$\pm$0.04\\
2002 Apr 7.294  & NTT          & SofI   & $K_s$ & 1800 & 0.7 &
	18.18$\pm$0.06\\
2002 Apr 7.381  & 1.54D        & DFOSC   & $R$ & 600 & 1.2 & 21.32$\pm$0.06\\
2002 Apr 8.280  & NTT          & SUSI2   & $V$ & 10$\times$60 & 0.8 & 
	22.49$\pm$0.02\\
2002 Apr 8.294  & NTT          & SUSI2   & $R$ & 10$\times$60 & 0.8 & 
	22.09$\pm$0.03\\
2002 Apr 8.310  & NTT          & SUSI2   & $I$ & 5$\times$60+3$\times$300
	& 0.8 & 21.44$\pm$0.02\\
2002 Apr 9.205  & NTT          & EMMI    & $R$ & 2$\times$300 & 1.0 & 
	22.23$\pm$0.04\\
2002 Apr 9.194  & NTT          & EMMI    & $V$ & 2$\times$300 & 1.0 & 
	22.76$\pm$0.06\\
2002 Apr 9.215  & NTT          & EMMI    & $I$ & 2$\times$300 & 1.0 & 
	21.70$\pm$0.06\\
2002 Apr 12.097 & {\it Yepun}  & FORS2   & $V$ & 3$\times$180 & 0.7 & 
	23.59$\pm$0.04\\
2002 Apr 12.107 & {\it Yepun}  & FORS2   & $R$ & 3$\times$180 & 0.7 & 
	23.09$\pm$0.04\\
2002 Apr 15.238 & {\it Melipal}& FORS1   & $V$ & 2$\times$180 & 0.7 & 
	23.99$\pm$0.07\\
2002 Apr 15.244 & {\it Melipal}& FORS1   & $R$ & 2$\times$180 & 0.6 & 
	23.36$\pm$0.04\\
2002 Apr 15.250 & {\it Melipal}& FORS1   & $I$ & 2$\times$180 & 0.6 & 
	22.53$\pm$0.06\\
2002 Apr 27.193 & {\it Antu}   & ISAAC  &$J_s$& 3600 &0.5 & 22.4$\pm$0.1\\
2002 May 8.193  & {\it Antu}   & ISAAC  &$J_s$& 3600 &0.6 & $>$22.7\\
\noalign{\smallskip}
\hline
\noalign{\smallskip}
\multicolumn{7}{l}{$^*$A preliminary analysis of this observation was
presented in Palazzi et al. (2002)}\\
\end{tabular}
\end{center}
\end{table*}

Optical images were bias-subtracted and flat-fielded with the standard
cleaning procedure. In some cases (especially at late epochs), frames
taken on the same night in the same band were summed together in order to
increase the signal-to-noise ratio. In Fig. 1 we report the R-band image
of the field of the GRB020405 counterpart obtained on 6 April 2002 with
the 1.54-metre Danish telescope plus DFOSC. 

Since the close environment of the afterglow is quite crowded (see Figs.
3 and 6), we chose standard Point Spread Function (PSF) fitting, rather
than simple aperture, photometry. We used the 
{\sl DAOPHOT II}~image data analysis package PSF-fitting algorithm 
(Stetson 1987) running within MIDAS\footnote{MIDAS (Munich Image Data 
Analysis System) is developed, distributed and maintained by ESO 
(European Southern Observatory) and is available at 
{\tt http://www.eso.org/projects/esomidas/}}.
A two-dimensional Gaussian profile with two free parameters (the half
width at half maxima along $x$ and $y$ coordinates of each frame) was
modeled on at least 5 unsaturated bright stars in each image. 
The errors associated with the measurements reported in Table 1 represent 
statistical uncertainties obtained with the standard PSF-fitting 
procedure.

The $BVRI$ zero-point calibration was performed by using several standard
fields (Landolt 1992) taken under photometric conditions at several ESO
telescopes. The single Gunn $i$ image (6 April 2002) was calibrated using
the $I$-band secondary standards, given that the widths, the reference
wavelengths and the flux density normalizations of the two filters are
very similar (Fukugita et al. 1995). However, in order to account for
small differences between the filters, we added in quadrature a 3\%
uncertainty to the $I$-band magnitude error obtained from the Gunn $i$
observation.

Concerning the single $U$-band measurement taken at WHT, no standard field
was acquired on the same night due to non-photometric conditions of the
sky. The calibration in this band was then performed on 10 May 2002 at the
1.0-metre JKT, located in the Canary Islands and equipped with a 
SITe2 CCD which has 2048$\times$2048 pixels and an image scale of 0.33 
pix$^{-1}$. Observations of the GRB field and of some Landolt (1992) 
fields were acquired in $U$ and $B$ to determine the CCD color term.

We then selected 6 stars of different brightness in the GRB020405 field
and we used them to determine the OT optical $BVRI$ magnitudes
(see also Simoncelli et al. 2002). Two additional stars were used for the
calibration in the $U$ band. All these stars are indicated in Fig. 1 and
their magnitudes are listed in Table 2. We find this magnitude calibration
to be accurate to within 3\%. 
We note that the $B$ and $R$ magnitudes of field stars as reported in the
USNO-A2.0 catalog\footnote{available at {\tt http://www.nofs.navy.mil/}}
differ by $\sim$0.6 mag with respect to our calibration, which is however
consistent with that of Bersier et al. (2003) and Price et al. (2003).

\begin{table*}
\caption[]{Optical magnitudes of selected GRB020405 field stars as
indicated in Fig. 1. Values are not reported for cases in which the 
stars are either saturated or too faint for reliable calibration}
\begin{center}   
\begin{tabular}{cccccc}
\noalign{\smallskip}
\hline
\noalign{\smallskip}
 Star & $U$ & $B$ & $V$ & $R$ & $I$ \\
\noalign{\smallskip}
\hline
\noalign{\smallskip}
1 & --- & 19.57$\pm$0.01 & 18.69$\pm$0.01 & 18.17$\pm$0.01 &
   17.55$\pm$0.01 \\ 
2 & --- & 19.84$\pm$0.01 & 19.01$\pm$0.01 & 18.48$\pm$0.01 &
   17.87$\pm$0.01 \\
3 & --- & 19.34$\pm$0.01 & 18.85$\pm$0.01 & 18.50$\pm$0.02 &
   18.12$\pm$0.01 \\
4 & --- & 20.09$\pm$0.02 & 19.49$\pm$0.01 & 19.14$\pm$0.02 &
   18.72$\pm$0.01 \\
5 & --- & 18.18$\pm$0.01 & 17.92$\pm$0.01 & 17.72$\pm$0.01 &
   17.45$\pm$0.01 \\
6 & --- & 20.45$\pm$0.02 & 19.35$\pm$0.01 & 18.67$\pm$0.01 &
   18.00$\pm$0.01 \\
A & 15.76$\pm$0.01 & 16.00$\pm$0.02 & --- & --- & --- \\ 
B & 18.35$\pm$0.03 & 17.74$\pm$0.02 & --- & --- & --- \\ 
\noalign{\smallskip}
\hline
\noalign{\smallskip}
\end{tabular}
\end{center}
\end{table*}

\begin{figure*}
\parbox{12.0cm}{
\psfig{file=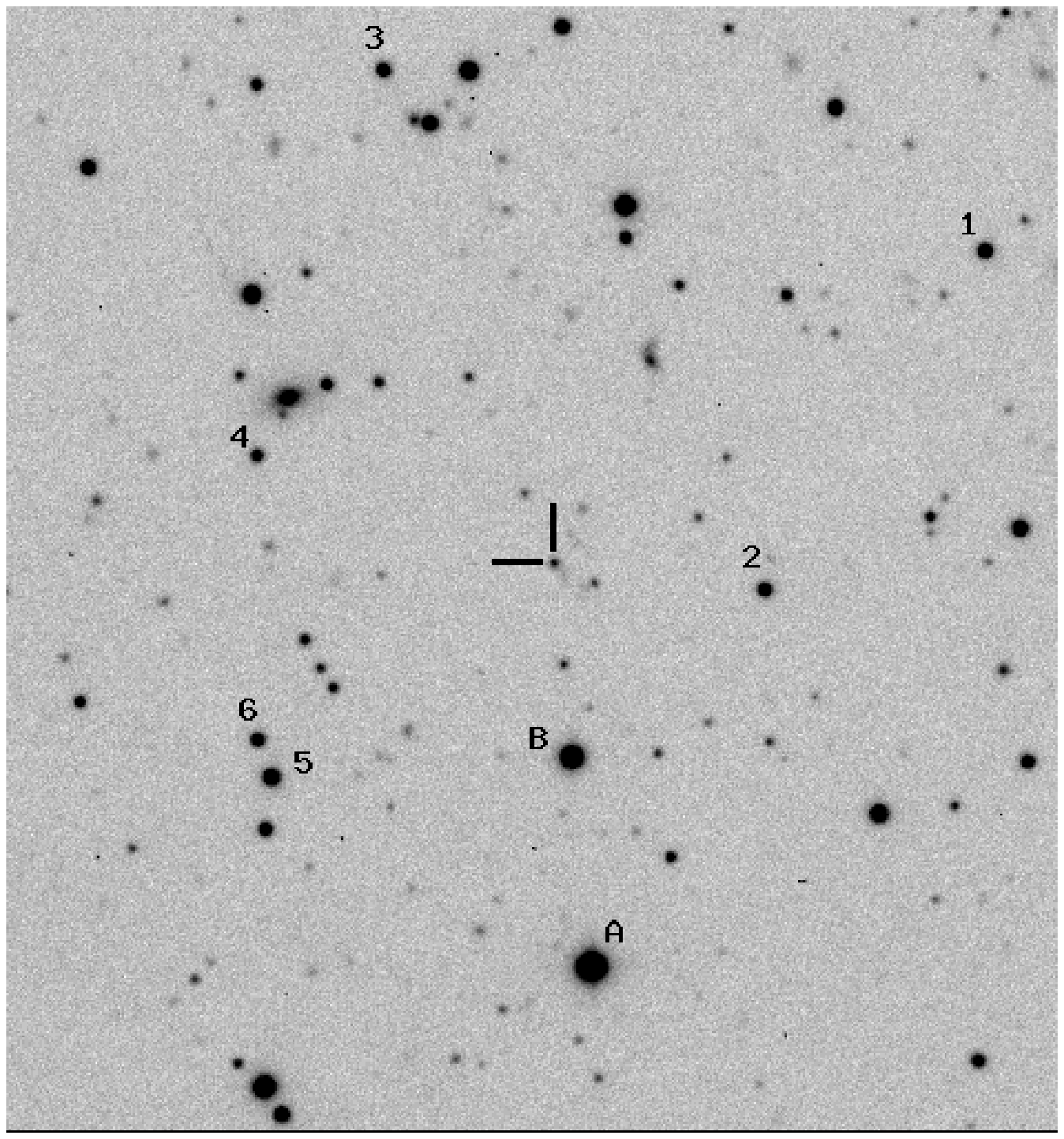,width=12cm}
}
\hspace{0.2cm}
\parbox{5.5cm}{
\vspace{9.2cm}
\caption{$R$-band image of the field of GRB020405 acquired with
the 1.54-metre Danish telescope plus DFOSC on 2002 April 6.136 UT at
ESO-La Silla.
The OT is at the centre of the image, indicated with the tick marks. 
North is at top, East is to the left; the field size is about
3$'$$\times$3$'$. Numbers and letters indicate the reference stars with
magnitudes reported in Table 2}
}
\end{figure*}

Public HST/WFPC2 observations of GRB020405 taken in three filters (F555W,
F702W and F814W) have been included in our analysis (see also Price et
al. 2003). All HST data were retrieved from the multimission
archive\footnote{see {\tt http://archive.stsci.edu/}} after ``On-The-Fly"
calibration had been performed. Dithered images were then combined and
cosmic-ray cleaned using the drizzle method (Fruchter \& Hook 2002). The
images were drizzled onto a final output grid with pixels half the size of
the native WPFC2 pixels using a {\tt pixfrac} parameter of 0.7.

The images acquired with filters F555W and F702W on August 23, 2002,
i.e., 140 days after the GRB, do not show substantial pointlike emission 
at the OT position; therefore they were used to evaluate the host galaxy 
contribution underlying the afterglow. 
This could not be done for the F814W data as no images were acquired with 
this filter during the last HST visit. Therefore, in order to estimate the 
magnitude of the host underlying the OT position in the F814W frames, we 
created a PSF using the {\sl Tiny Tim} software (Krist 1993). This PSF was 
subsampled and convolved to mimic the effect of the drizzle. The PSF was 
then subtracted from the position of the OT in the June 2002 F814W HST 
images to the level where the remaining surface brightness was entirely 
due to the host.
These PSF-subtracted images were then used as a template to determine and 
subtract the host galaxy contribution in all previous (space- and 
ground-based) observations made in that band.
The accuracy of this method was checked by applying it to the June 2002 
images acquired with filters F555W and F702W, and by comparing the results
with those obtained using the August 2002 data.
We found that the host magnitudes measured at the OT position with the two 
methods (i.e., direct host contribution measurement on one hand, and PSF 
modeling plus subtraction on the other) coincide within the uncertainties.
This made us confident that the host contribution, as derived with
the PSF-subtraction method, was correct and could be applied to the F814W
images.

OT magnitudes were then measured on each available HST image with standard
aperture photometry, and were calibrated to the Vega system using the
published zeropoints for the WFPC2 (Biretta et al. 2002). 
We assumed that the conversion from the WFPC2 to the Johnson-Cousins 
system did not induce significant errors on the photometry, as also stated 
in Fukugita et al. (1995). Indeed, the resulting OT magnitudes, reported 
in Table 3, are consistent with the flux values independently reported by 
Price et al. (2003). The only discrepancy
is found in the F814W filter data and arises from the fact that those
authors assumed no OT contribution in the June 2002 HST image acquired
with that filter, so that was used directly as the template for the host
galaxy subtraction. In fact, if we subtract, for that HST filter, the June
2002 magnitude of the OT plus underlying host from those obtained in the
previous visits, we obtain flux density values consistent with those
reported in Table 3 of Price et al. (2003).

The reduction of the NIR images was performed with IRAF and the STSDAS
packages\footnote{IRAF is the Image Analysis and Reduction Facility
made available to the astronomical community by the National Optical
Astronomy Observatories, which are operated by AURA, Inc., under
contract with the U.S. National Science Foundation.
STSDAS is distributed by the Space
Telescope Science Institute, which is operated by the Association of
Universities for Research in Astronomy (AURA), Inc., under NASA contract
NAS 5--26555.}. Each image was reduced by first subtracting a mean 
sky, obtained from the median of a number of images acquired just before
and after each processed frame.
Before the frames were used for sky subtraction, stars in them
were eliminated by a background interpolation algorithm ({\it imedit})
combined with an automatic ``star finder" ({\it daofind}).
Then, a differential dome flatfield correction was applied;
next, the telescope dithering was measured from the offsets of field 
objects in each image and the images were averaged together using 
inter-pixel shifts.
Also for the NIR images we used PSF-fitting photometry following the same 
procedure adopted in the optical. We calibrated the NIR 
photometry using stars selected from the NICMOS Standards List (Persson 
et al. 1998) and determined the NIR zero-point coefficients for each 
observation. 
The standard stars were observed in five positions on the detector,
and their images were reduced in the same way as those of the GRB field.
Formal photometric accuracy based only on the standard stars observations 
is typically better than 3\%. Our calibration is fully consistent with 
the results of the 2MASS 
survey\footnote{available at {\tt http://www.ipac.caltech.edu/2mass/}}
(Skrutskie et al. 1997) for the GRB020405 field.

Next we evaluated the Galactic absorption in the optical and NIR bands
along the direction of GRB020405 using the Galactic dust infrared maps
by Schlegel et al. (1998); from these data we obtained a color excess
$E(B-V)$ = 0.055. By applying the relation by Cardelli et al. (1989), we 
derived $A_U$ = 0.27, $A_B$ = 0.22, $A_V$ = 0.17, $A_R$ = 0.14, 
$A_I$ = 0.10, $A_J$ = 0.05, $A_H$ = 0.03 and $A_{K_s}$ = 0.02.

\subsection{Optical spectroscopy}

Two series of spectra with total exposure times of 30 minutes and 60 
minutes were obtained on 2002 April 6.213 and 7.310 UT, respectively, 
with VLT-{\it Melipal}+FORS1.
The first series was acquired using Grism 150I plus order separator 
OG590, which avoids overlapping of spectral orders over a given 
wavelength; this limited the spectral range to 6000-9000 \AA.
For the second series the Grism 300V with no order separator was 
used: this allowed a wider spectral coverage (3500-9000 \AA) and a 
spectral resolution higher by about a factor of two.
The slit width was 1$''$ for both observations and these setups 
secured final dispersions of 5.5 \AA~pix$^{-1}$ for the first series 
and of 2.6 \AA~pix$^{-1}$ for the second. 
In the second spectroscopic pointing the slit was rotated by about
40$^{\circ}$ with respect to the N-S direction in order to include
both the OT and the galaxy located $\sim$2$''$ southwest of it
(Hjorth et al. 2002).

The spectra, after correction for flat-field and bias, were background
subtracted and optimally extracted (Horne 1986) using IRAF.
He-Ne-Ar and Hg-Cd lamps were used for wavelength calibration; both 
spectroscopic runs were then flux-calibrated by using the spectroscopic 
standards LTT 6248 and Hiltner 600 (Hamuy et al. 1992, 1994) for the 
April 6 and 7 observations, respectively. Finally, spectra taken within 
the same night were stacked together to increase the S/N ratio.
The correctness of the wavelength and flux calibrations was checked
against the position of night sky lines and the photometric data collected
around the epoch at which the spectra were acquired, respectively.
The typical errors were 0.3 \AA~for the wavelength calibration and 10\%
for the flux calibration.

\begin{table}
\caption[]{HST-WFPC2 optical magnitudes of the GRB020405 OT. The values 
are corrected for the host galaxy contribution, but not for the Galactic
absorption}
\begin{center}
\begin{tabular}{lcc}
\noalign{\smallskip}
\hline
\noalign{\smallskip}
\multicolumn{1}{c}{Mid-exposure date (UT)} & Filter & Magnitude \\
\noalign{\smallskip}
\hline
\noalign{\smallskip}

2002 Apr 24.272 & F555W & 25.05$\pm$0.04 \\
2002 May 5.631  & F555W & 25.83$\pm$0.05 \\
2002 Jun 2.711  & F555W & 27.25$\pm$0.15 \\
2002 Aug 23.246 & F555W & $>$27.5 \\

 & & \\

2002 Apr 28.452 & F702W & 23.99$\pm$0.02 \\
2002 May 1.620  & F702W & 24.31$\pm$0.02 \\
2002 May 3.625  & F702W & 24.38$\pm$0.02 \\
2002 Jun 1.614  & F702W & 25.91$\pm$0.12 \\
2002 Aug 23.451 & F702W & $>$26.7 \\

 & & \\

2002 Apr 26.275 & F814W & 23.13$\pm$0.01 \\
2002 May 1.486  & F814W & 23.45$\pm$0.02 \\
2002 Jun 9.560  & F814W & 25.42$\pm$0.15 \\

\noalign{\smallskip}
\hline
\noalign{\smallskip}
\end{tabular}
\end{center}
\end{table}

\subsection{Optical polarimetry}

Linear polarimetry in the $R$ band was accomplished on 2002 April 
6.236 UT at VLT-{\it Melipal}+FORS1 under an average seeing of 
0$\farcs$8. Data were obtained by using a Wollaston prism and a 
half-wavelength, phase-retarder plate. 
The Wollaston prism separates the incident light into an ordinary and an 
extraordinary component, while the
phase-retarder plate determines which of the Stokes parameters
is measured ($U$ or $Q$). For each image, a mask producing 22$''$-wide
parallel strips was used to avoid overlap of the ordinary and
extraordinary components. The observation therefore consisted of four
exposures centered at the position of the OT, with the phase-retarder
plate at 0$^{\circ}$, 22$\fdeg$5, 45$^{\circ}$ and 67$\fdeg$5. Each angle 
was imaged with an exposure time of 300 s, for a total exposure time of 
1200 s over the four angles.
Image reduction and analysis was performed as described in Sect. 2.1
for the optical photometry. 
No significant variation among the values of the $Q$ and $U$ parameters
of individual field stars was noted, therefore we computed 
the average values of
these quantities and subtracted them from the corresponding parameters 
of the OT to remove the instrumental and (local) interstellar polarization.
We also checked that the polarization of the selected field stars did 
not systematically vary with their position on the CCD or with their 
magnitude. The polarization angle $\theta$ was calibrated by
using the polarized standard Vela1\_95 (Szeifert 2001). The 
unpolarized standard WD 1620$-$315 (Szeifert 2001) was used to check the 
instrumental polarization, which was found to be negligible.

\section{Results}

The optical (see Fig. 1) and NIR images show a point-like source at 
the OT position indicated by Price et al. (2002a, 2003). The transient 
behaviour observed in both spectral ranges (see Table 1) confirms the 
afterglow nature of the source.

\subsection{Light curves}

\begin{figure*}
\parbox{14cm}{
\psfig{file=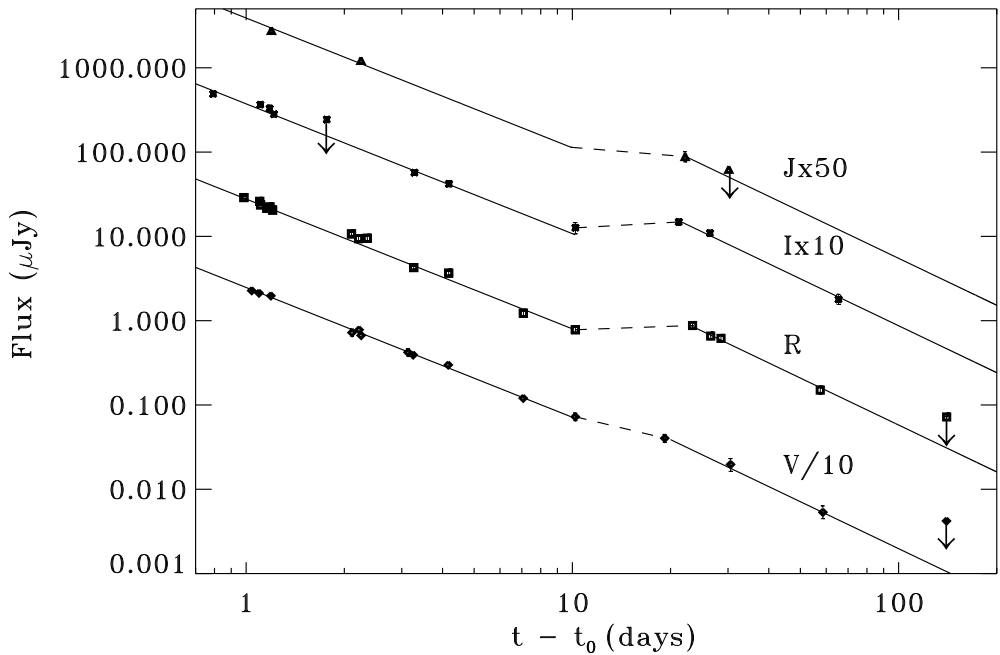,width=13.5cm}
}
\hspace{-0.5cm}
\parbox{4.5cm}{
\vspace{2.5cm}
\caption{$VRIJ$ light curves of the GRB020405 afterglow. Different symbol
styles indicate different bands. Data are corrected for the underlying
host galaxy contribution and for Galactic absorption.
The light curves were rescaled in flux for clarity using the factors
indicated in the figure.
A flattening between day 10 and 20 after the GRB, more evident in 
the $RIJ$ bands, is detected.
The light curves are best fitted using a power law with 
$\alpha$ = 1.54 (up to 10 days after the GRB) and a power law with 
$\alpha'$ = 1.85 (after day 20)}
}
\end{figure*}

In the analysis of the optical and NIR light curves of the GRB020405 we
added to our data set the points published by Covino et al. (2003),
whose photometry can be connected with ours.
We first considered the ground-based data. These may be contaminated by
the underlying host galaxy complex. Thus, in order to 
correct for its contribution to the $VRI$ magnitudes of the afterglow, 
we followed two approaches: (i) we fit the light
curves with a single power law plus constant ($F \propto t^{-\alpha}$ +
$const$); (ii) we measured, on the latest available (OT-subtracted) HST 
images, the contribution of the host galaxy complex within an aperture 
radius matching the ground-based telescopes PSFs and subtracted it from
our photometric measurements. The two methods gave results which are 
consistent within the errors, and showed that the host contribution to 
the OT luminosity was of the order of few percent (and thus negligible if 
compared with the uncertainties) in the first days of observation. 

Concerning the $J$-band data, in the VLT-ISAAC image of 8 May 2002
we did not detect a point-like source at the position of the afterglow, 
despite the depth and good resolution of the image. Therefore we evaluated 
the host galaxy contribution in the $J$ band by using aperture photometry 
centered at the OT position and with radius equal to the FWHM of the image 
PSF. This can be conservatively considered as an upper limit to the OT 
flux.

No late-time HST or ground-based data are available for other
optical or NIR filters. Therefore, no host-subtraction could be done for 
bands different than $VRIJ$. Since the host contribution could be evaluated 
only for these four bands, only the corresponding light curves are reported 
in Fig. 2.

The host-subtracted $VRI$ light curves between 1 and 10 days after the GRB
decay according to single power laws whose temporal indices are not
significantly different from filter to filter. Their average value is
$\alpha$ = 1.54$\pm$0.06. The $B$ light curve, which consists of few,
early epoch points, is fitted by the same temporal power law. By modeling
the light curves using a smoothly broken power law as described by
Beuermann et al. (1999) we obtain a worse fit (i.e., with a larger 
reduced $\chi^2$), indicating that no temporal break is apparent in our 
optical data set. 

While the $J$-band light curve is reasonably
fitted with a single power law with index compatible with that measured
for the optical data, the $H$ and $K_s$ data appear to decay
with a shallower slope, $\alpha \sim$ 1.3. Considering only
the NIR data acquired in the first two days (which are presumably least 
contaminated by the host galaxy), we get an average decay slope
$\alpha_{\rm NIR}$ = 1.28$\pm$0.03, substantially 
shallower than the one measured in the optical. 

As already noted for this (Price et al. 2003) and for other, well sampled, 
afterglows (e.g., GRB010222: Masetti et al. 2001), a single or 
smoothly broken power-law fit does not formally give an acceptable fit to 
the $VRI$ data. However, if we quadratically add a constant 
uncertainty of less than 3\% to all $VRI$ points in our data set, 
the single power-law model gives an acceptable fit (with $\chi^2_\nu \sim$ 
1) for the same $\alpha$ obtained before.
This formal discrepancy between the data and the model might be due 
to instrumental differences (e.g., the use of different detectors and 
different telescopes) or to intrinsic short time scale variations of the 
afterglow light curve (see e.g., Masetti et al. 2000; Holland et al. 2002; 
Jakobsson et al.  2002; Holland et al. 2003), possibly due to 
inhomogeneities of the circumburst medium (Wang \& Loeb 2000).

Between day 10 and 20 after the GRB, the OT decay flattens in all bands
but more evidently in the $RIJ$ bands, thus producing the red 
bump first noted by Price et al. (2003); thereafter, the flux decreases 
at a faster rate (Fig. 2).
We fitted the data after day 20 to a single power law with index nearly 
independent of the wavelength and of average value $\alpha'$ = 1.85$\pm$0.15.
Interestingly, the 8.45 GHz light curve of the radio afterglow
(Berger et al. 2003) closely resembles this late-time optical 
behaviour.

\begin{figure*}[t!]
\psfig{file=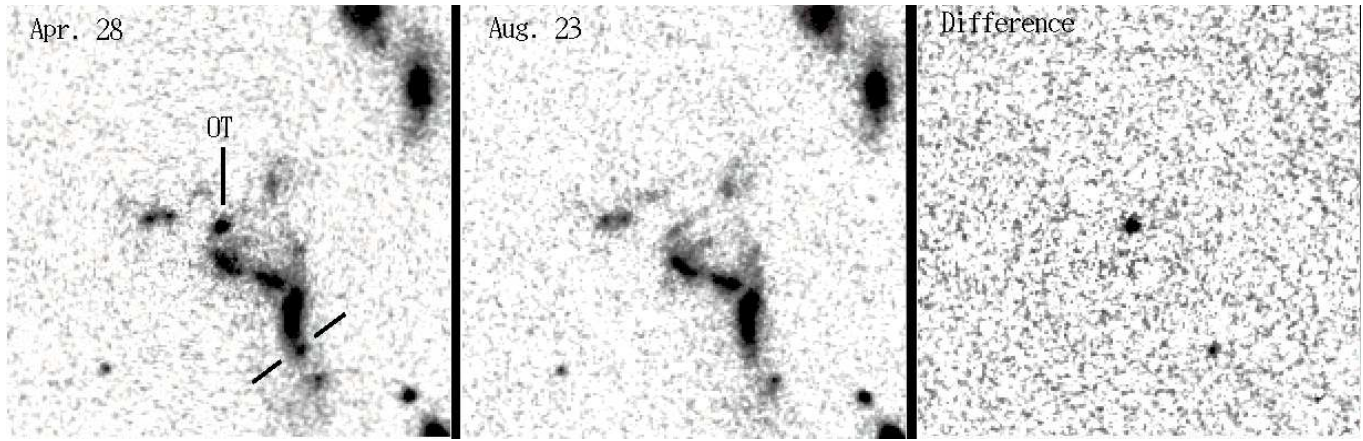,width=18cm}
\caption{HST-WFPC2 images of the OT field acquired with the F702W
filter on April 28 (left panel) and August 23 (central panel), along with
their difference (right panel). Besides the OT, located at the centre of
the image, a second transient object (also indicated with tick marks in
the April 28 image) is apparent $\sim$3$''$ southwest of the OT itself.
The images are about 10$''$$\times$10$''$ in size; North is at top, East
is to the left}
\end{figure*}

\begin{figure}[h!]
\hspace{-.3cm}
\psfig{file=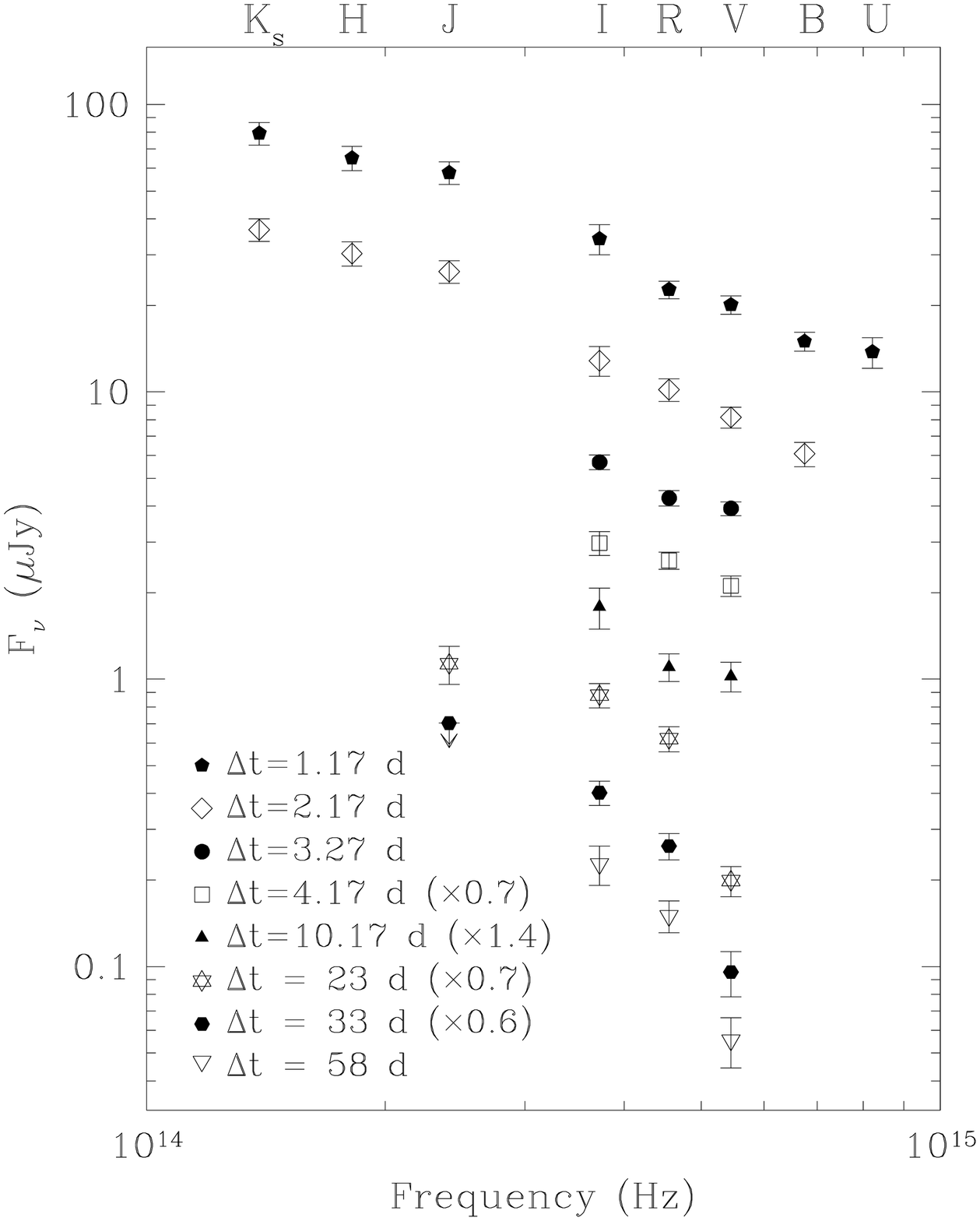,width=9.2cm}
\vspace{-1.2cm}
\caption{Optical-NIR broad-band spectra of the
GRB020405 afterglow at various epochs between $\sim$1 and $\sim$60 days
after the GRB. Different symbol styles refer to different epochs.
When needed, the broad-band spectra were rescaled in flux 
for clarity using the factors indicated in the figure.
A spectral change located at (2.5$\pm$0.6)$\times$10$^{14}$ Hz is apparent 
in the first two epochs.
Assuming a broken power-law shape, optical and NIR spectral slopes are
all consistent with being 1.3 and 0.65, respectively, during the first
$\sim$10 days. In the last three epochs, the optical slope becomes 3.5, 
thus significantly steeper, and a pronounced spectral curvature is
apparent between the optical and NIR ranges}
\end{figure}

Finally we note that the HST images in the filters F702W and F814W
reveal the presence of a transient source located in the southern
outskirts of galaxy `1' (as defined in Sect. 3.3 and Fig. 6), at
coordinates (J2000)  RA = 13$^{\rm h}$ 58$^{\rm m}$ 03$\fs$01; Dec =
$-$31$^{\circ}$ 22$'$ 24$\farcs$7 (J2000), with a conservative uncertainty
of 0$\farcs$1 on both. In Fig. 3 the position of this second transient is
shown. This object faded between the end of April and the beginning of
June 2002, and is not detected in the August 2002 HST visit. It is also
not detected in any of the F555W HST images. A more detailed description
of this source will be given in a forthcoming paper.

\subsection{Optical-NIR spectral flux distribution}

We have constructed optical-NIR broad-band spectra at 8 epochs 
spanning from $\sim$1 to $\sim$60 days after the GRB and for which we have 
sufficient optical and/or NIR coverage (Fig. 4). The data points were 
corrected for the Galactic absorption and converted into flux densities
using the tables by Fukugita et al.
(1995) for the optical and by Bersanelli et al. (1991) for the NIR.
The spectra of the first two epochs were not corrected for the host
contribution because this is known for the $VRIJ$ bands only; however,
given this was quite modest at those epochs (see Sect. 3.1),
we simply added a 5\% error in quadrature to the uncertainties on the
optical-NIR flux densities. 
The subsequent spectra were also corrected for the host galaxy
contribution; the errors on the points include the uncertainty on the host
photometry.

For the early epoch (within $\sim$10 days after the GRB) 
broad-band spectra, in the cases in which the data in different bands were 
not simultaneous, the flux density was interpolated to the reference epoch
assuming the best-fit power-law decay (see Sect. 3.1). 
In the last three broad-band spectra the interpolations to a common epoch
were computed using the power-law decay slope ($\alpha'$ = 1.85)
determined from observations acquired after day 20 from the GRB.

Single power law ($F_\nu \propto \nu^{-\beta}$) fits of the first two
broad-band spectra are unacceptable due to a significant deviation of the
J-band points. Since we can firmly exclude instrumental effects or
systematic errors in the $J$-band calibration, we fitted the spectra with
broken power laws. The break is found to be located at
(2.5$\pm$0.6)$\times$10$^{14}$ Hz, i.e., in the $J$ band, and the 
slopes are $\beta_{\rm NIR}$ = 0.65$\pm$0.2 and $\beta_{\rm opt}$ = 
1.3$\pm$0.2, with no significant variability of these parameters between 
the two epochs. The optical slope agrees well, within the uncertainties, 
with those determined by Price et al. (2003) and Bersier et al. (2003). 

The broad-band optical spectra of April 8.3, 9.2 and 15.2 UT were well
fitted with single power laws, with spectral slopes 1.0$\pm$0.2,
0.9$\pm$0.3 and 1.35$\pm$0.5, respectively; these are consistent with the
optical slope measured in the first two epochs. 
The OT emission during the ``red bump", measured 23, 33, and 58 days
after the GRB, is fitted by a power law of average spectral index $\beta$
= 3.5$\pm$0.5, in agreement with the findings of Price et al. (2003) and
substantially steeper than at the earlier epochs. A remarkable
NIR-to-optical spectral curvature is observed.

The optical and NIR colors of the OT of GRB020405 during the first 10 
days of monitoring after the high-energy event fall in the loci populated 
by GRB afterglows in the color-color diagrams as illustrated by \v{S}imon 
et al. (2001) and by Gorosabel et al. (2002).

\subsection{Optical spectra}

\begin{figure*}[t!]
\psfig{file=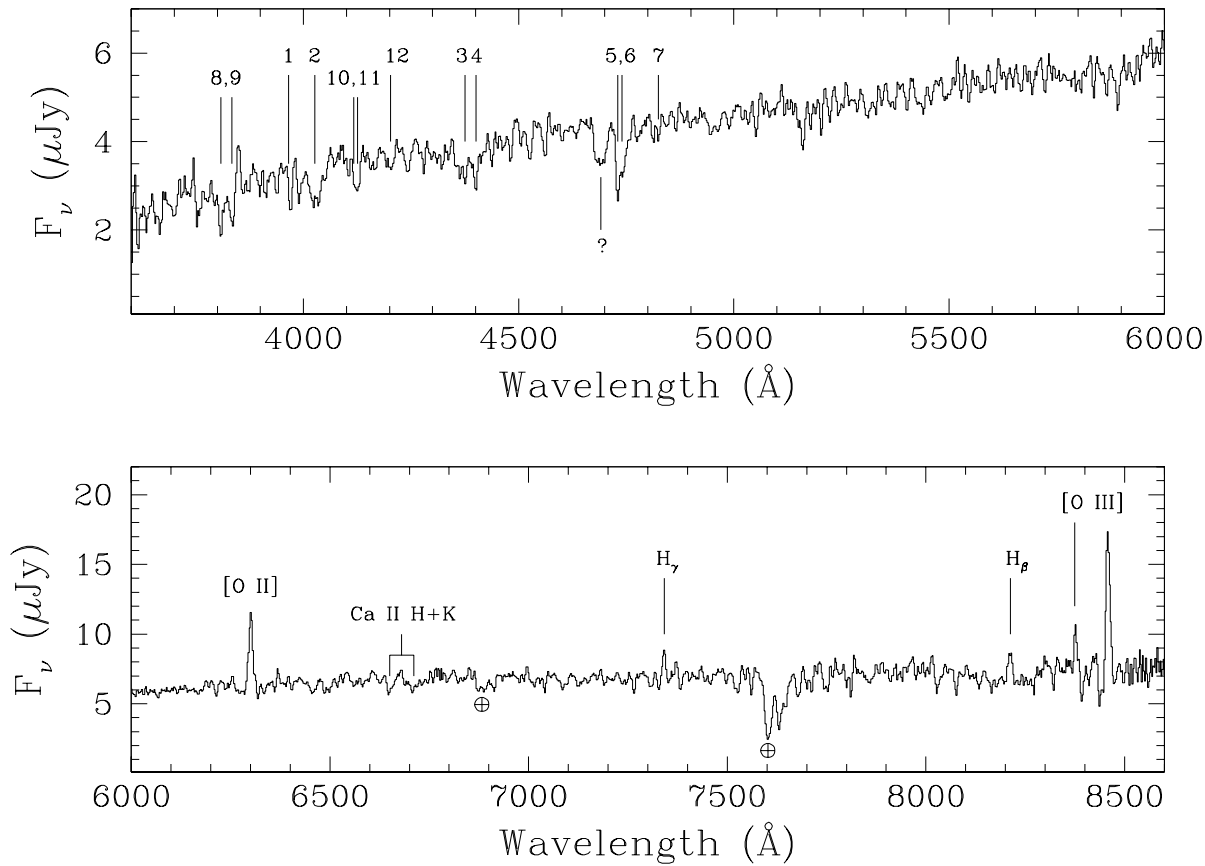,width=18cm,angle=-90}
\vspace{-0.8cm}
\caption{Spectrum of the OT of GRB020405 obtained with VLT-{\it Melipal} 
plus FORS1 on 7 April 2002. The spectrum was smoothed with a Gaussian
filter with 
$\sigma$ = 3 \AA~(i.e., comparable with the spectral resolution) and 
corrected for foreground Galactic absorption assuming $E(B-V)$ = 0.055. 
For clarity, we split it into two panels. Numbers mark the positions of the 
identified absorption lines as listed in Table 4, while the question mark
indicates an unidentified line. Telluric absorption features at 6870
\AA~and 7600 \AA~are indicated with the symbol $\oplus$}
\end{figure*}

\begin{table*}[t!]
\caption[]{List of emission and absorption lines identified in the VLT 
optical spectra of the GRB020405 OT acquired on April 6 and 7. 
The numbers in the leftmost column refer to the identifications shown in 
Fig. 5. The error on line positions is assumed to be $\pm$6 \AA~for the 
spectrum of April 6 and $\pm$3 \AA~for that of April 7, i.e., 
comparable with the corresponding spectral resolution (see text). 
For absorption lines we report the rest-frame EW, and for emission lines
their intensity}

\begin{center}
\begin{tabular}{ccclclr}
\noalign{\smallskip}
\hline
\noalign{\smallskip}
\multicolumn{1}{c}{Line} & Observed & Rest frame & 
\multicolumn{1}{c}{Transition} & Redshift & 
\multicolumn{1}{c}{EW$_{\rm rest}$} & \multicolumn{1}{c}{Flux}\\
\multicolumn{1}{c}{number} & wavelength (\AA) & wavelength (\AA) & & &
\multicolumn{1}{c}{(\AA)} & ($\times$10$^{-17}$ erg cm$^{-2}$ s$^{-1}$)\\
\noalign{\smallskip}
\hline
\noalign{\smallskip}
\multicolumn{7}{c}{6 April 2002} \\
\noalign{\smallskip}
\hline
\noalign{\smallskip}
& 6302 & 3727 & [O {\sc ii}]  & 0.6908$\pm$0.0016 & & 9.5$\pm$1.5 \\
& 6647 & 3935 & Ca {\sc ii} K & 0.6893$\pm$0.0015 & 1.2$\pm$0.3 & \\
& 6710 & 3969 & Ca {\sc ii} H & 0.6905$\pm$0.0015 & 1.2$\pm$0.3 & \\
& 8227 & 4861 & H$_\beta$     & 0.6923$\pm$0.0012 & & 4.0$\pm$0.5 \\
& 8386 & 4959 & [O {\sc iii}] & 0.6910$\pm$0.0012 & & 4.0$\pm$0.5 \\
& 8472 & 5007 & [O {\sc iii}] & 0.6920$\pm$0.0012 & & 11.5$\pm$1.5 \\
\noalign{\smallskip}
\multicolumn{3}{r}{Weighted mean} & ........................... &
0.6910$\pm$0.0014 & & \\
\noalign{\smallskip}
\hline
\noalign{\smallskip}
\multicolumn{7}{c}{7 April 2002} \\
\noalign{\smallskip}
\hline
\noalign{\smallskip}
1 & 3965 & 2344 & Fe {\sc ii}   & 0.6912$\pm$0.0013 &    1.5$\pm$0.3 & \\
2 & 4026 & 2374+2383 & Fe {\sc ii}$^*$& 0.6916$\pm$0.0013 &  6.0$\pm$1.2 & \\
3 & 4375 & 2587 & Fe {\sc ii}   & 0.6915$\pm$0.0012 &    0.8$\pm$0.3 & \\
4 & 4401 & 2600 & Fe {\sc ii}   & 0.6889$\pm$0.0012 &    1.1$\pm$0.3 & \\
5 & 4730 & 2796 & Mg {\sc ii}   & 0.6913$\pm$0.0011 &    2.7$\pm$0.4 & \\
6 & 4740 & 2803 & Mg {\sc ii}   & 0.6909$\pm$0.0011 &    1.2$\pm$0.3 & \\
7 & 4825 & 2853 & Mg {\sc i}    & 0.6913$\pm$0.0011 &    0.4$\pm$0.2$^{**}$ &\\
  & 6302 & 3727 & [O {\sc ii}]  & 0.6908$\pm$0.0008 & & 8.5$\pm$1.0  \\
  & 6650 & 3935 & Ca {\sc ii} K & 0.6900$\pm$0.0008 &    1.2$\pm$0.3 & \\
  & 6711 & 3969 & Ca {\sc ii} H & 0.6905$\pm$0.0008 &    1.5$\pm$0.4 & \\
  & 7342 & 4341 & H$_\gamma$    & 0.6915$\pm$0.0007 & & 1.3$\pm$0.5 \\
  & 8213 & 4861 & H$_\beta$     & 0.6895$\pm$0.0006 & & 2.5$\pm$1.0 \\
  & 8375 & 4959 & [O {\sc iii}] & 0.6888$\pm$0.0006 & & 3.0$\pm$1.0 \\
  & 8460 & 5007 & [O {\sc iii}] & 0.6897$\pm$0.0006 & & 11.0$\pm$2.0 \\
\noalign{\smallskip}
\multicolumn{3}{r}{Weighted mean} & ........................... &
0.6905$\pm$0.0010 & & \\
\noalign{\smallskip}
\hline
\noalign{\smallskip}
 8 & 3808 & 2587 & Fe {\sc ii} & 0.4720$\pm$0.0012 & 2.4$\pm$0.4 & \\
 9 & 3834 & 2600 & Fe {\sc ii} & 0.4714$\pm$0.0012 & 2.5$\pm$0.4 & \\
10 & 4117 & 2796 & Mg {\sc ii} & 0.4723$\pm$0.0011 & 1.1$\pm$0.3 & \\
11 & 4126 & 2803 & Mg {\sc ii} & 0.4717$\pm$0.0011 & 1.1$\pm$0.3 & \\
12 & 4202 & 2853 & Mg {\sc i}  & 0.4729$\pm$0.0011 & 0.4$\pm$0.3$^{**}$ & \\
\noalign{\smallskip}
\multicolumn{3}{r}{Weighted mean} & ........................... &
0.4720$\pm$0.0011 & & \\
\noalign{\smallskip}
\hline
\hline
\noalign{\smallskip}
\multicolumn{7}{l}{$^*$Blend of two Fe {\sc ii} lines. The restframe
centroid is roughly estimated to be at 2380 \AA}\\
\multicolumn{7}{l}{$^{**}$This identification is only tentative;
it is however included in the list as it coincides with the redshift
of this absorption system}\\
\end{tabular}
\end{center}
\end{table*}

The continuum slopes measured for the VLT spectra of April 6 and 7 are 
consistent with the results of our broad-band spectral analysis reported
in the previous Subsection.

In the OT spectra of April 6 and 7 we detected a number of emission lines
which we identified with the forbidden transitions of [O {\sc ii}]
$\lambda$3727, [O {\sc iii}] $\lambda\lambda$4959,5007, and with the
H$_\beta$ line at an average redshift of 0.6908$\pm$0.0017.
In the spectrum of April 7 (Fig. 5) we also possibly detect H$_\gamma$ 
in emission.
Our redshift measurement is confirmed by the detection of several 
absorption features blueward of 5000 \AA~in the April 7 spectrum.
These absorptions are due to rest-frame ultraviolet (UV) metallic lines 
usually present in OT spectra and, more generally, in the spectra of 
high-$z$ objects (see, e.g., Savaglio et al. 2003 and references 
therein). In this wavelength range we also detect an
absorption system at a lower average redshift $z$ = 0.4720$\pm$0.0011. 

Our line fitting, performed with the SPLOT task within IRAF, assumes a
Gaussian profile for both emission and absorption lines. A conservative
error of 6 \AA~and 3 \AA, comparable with the spectral resolution, is 
associated with line wavelength measurements in the spectra
of April 6 and 7, respectively. In Table 4 are reported, for each line 
detected in the two spectroscopic runs, the identification along with the 
observed and rest-frame wavelengths and the associated redshift. 
No significant
difference in the emission and absorption line redshifts is found in
either observation for the features at $z$ = 0.691. In Table 4 we also
list the equivalent widths (EWs) of the identified lines computed in the
absorber rest frame, i.e., dividing the measured value by the factor
(1+$z$). The errors on the EWs are computed by assuming for the spectral
continuum in proximity of each line the values corresponding to its 
1-$\sigma$ lower and upper bounds.
We did not find any significant variation in the emission line fluxes
between the two nights, as expected if the line production 
sites are gas-rich and star-forming regions of the GRB host galaxy.
The EWs of the only absorption lines detected in both nights (those of 
Ca {\sc ii}) do not vary significantly.
We also detected a broad absorption feature (the observer's frame EW is
5.6$\pm$0.5 \AA) at 4691 \AA~which cannot be identified with any known 
line among those typically detected in OT spectra at either of the two 
detected redshifts.
This feature is indicated with a question mark in Fig. 5.

The slit orientation used for the spectroscopic 
observations acquired on April 7 also allowed us to acquire spectra 
of the nearby galaxy first noted by Hjorth et al. (2002). Fig. 6 
reports a sketch of the setup with which the VLT spectrum of April 7 was 
obtained: the 1$''$-wide slit was rotated by 40$^\circ$ towards East with 
respect to the N-S direction in order to acquire also the spectrum of the
galaxy at 2$''$ southwest of the OT, marked as `1' in Fig. 6 (see also
Hjorth et al. 2002). The slit also included a galaxy
(indicated as `2' in the same Figure) located $\sim$6$''$ southwest of the
OT. The spectra of these two galaxies are reported in Fig. 7.

\begin{figure}[t!]
\psfig{file=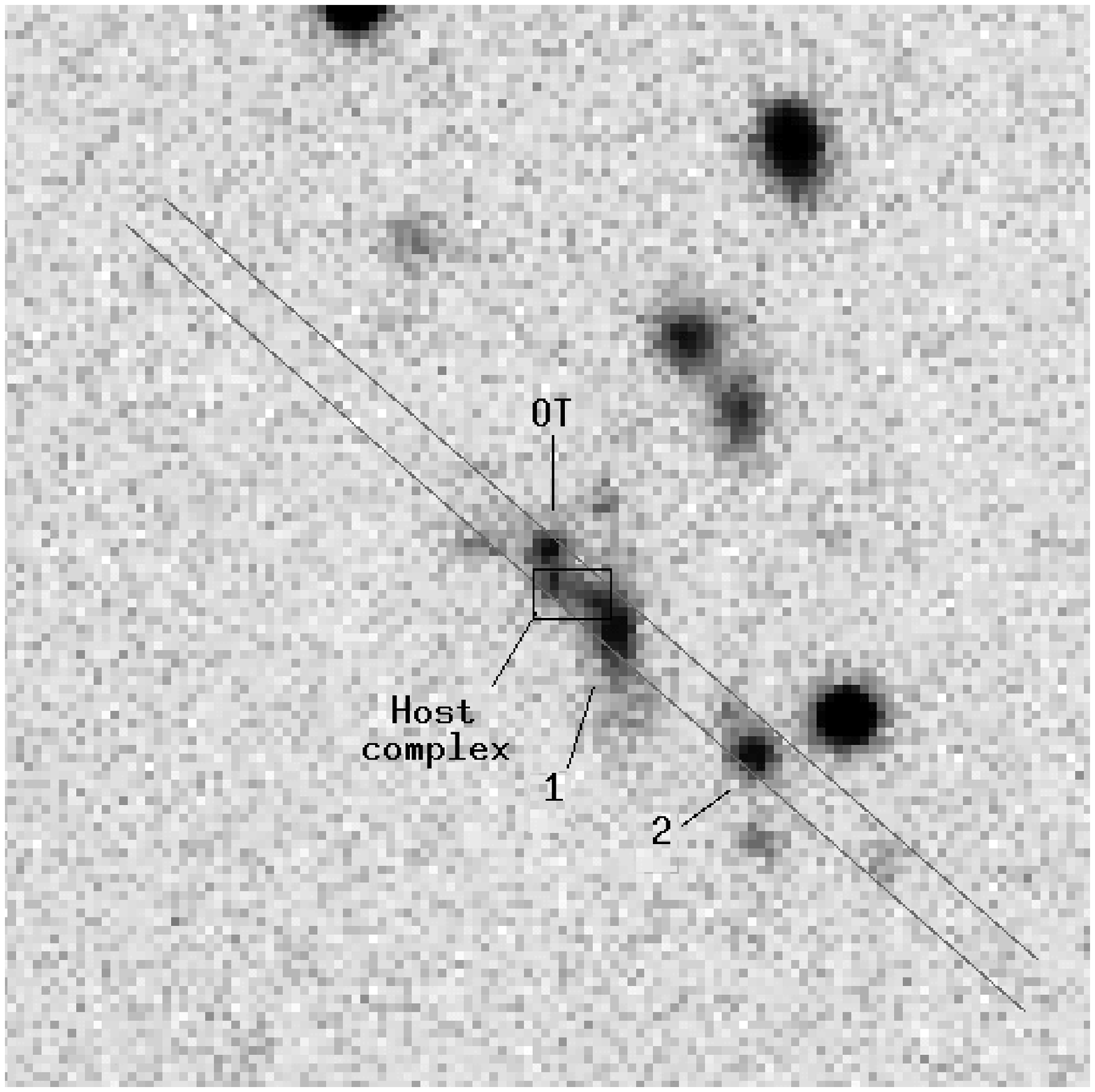,width=8.8cm}
\caption{VLT-{\it Melipal}(+FORS1) $R$-band image of the GRB020405 field
acquired on 15 April 2002. The host galaxy of the GRB and galaxies `1'
and `2' described in the text are indicated. The field size is about 
25$''$$\times$25$''$; North is at top, East is to the left. 
The diagonal solid lines indicate the slit position and width of the VLT 
spectrum obtained on 7 April 2002}
\end{figure}

\begin{figure}
\hspace{-.3cm}
\psfig{file=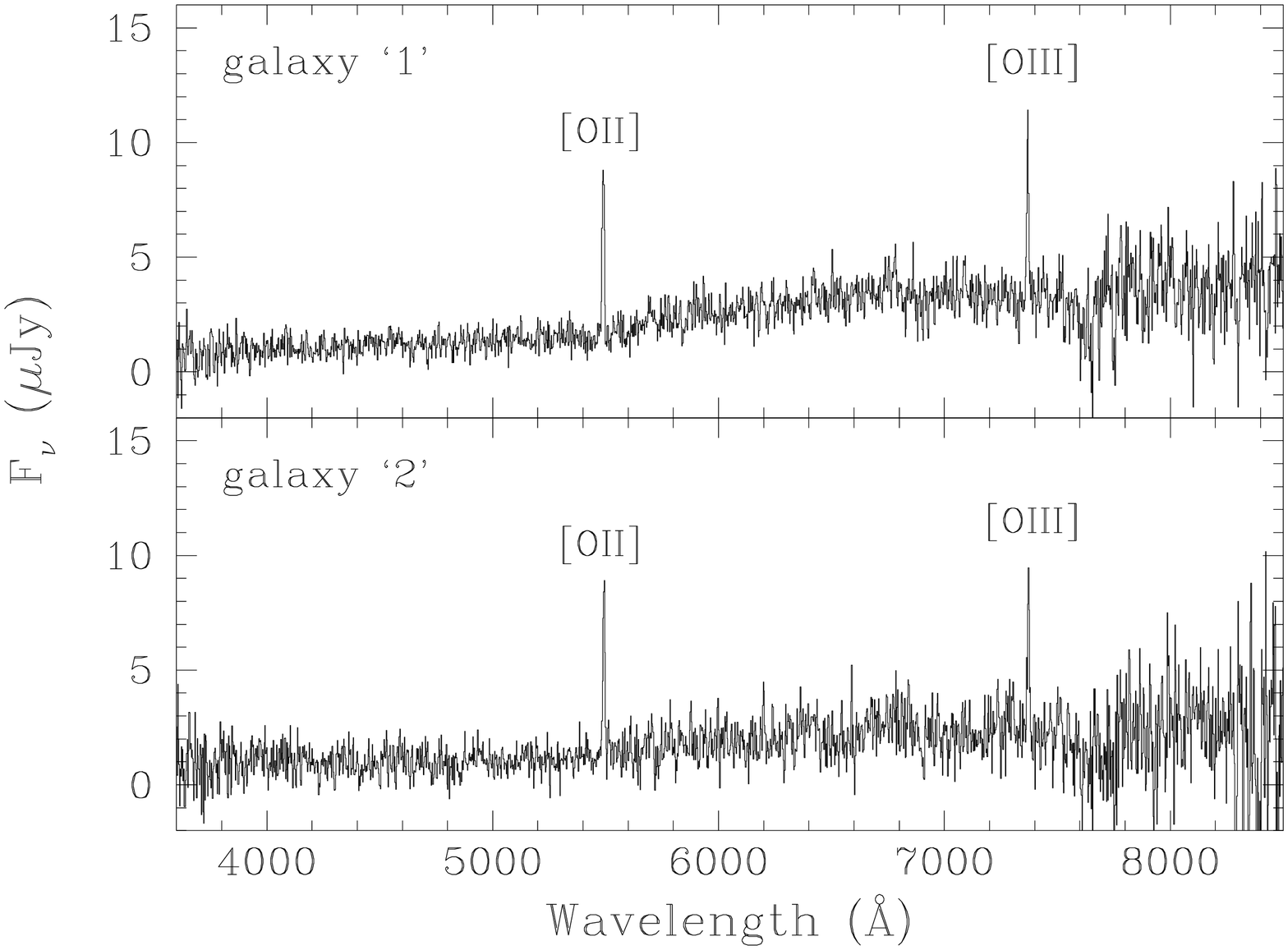,width=9.9cm,angle=-90}
\vspace{-.7cm}
\caption{Spectra of galaxies `1' {\it (upper panel)} and `2' 
{\it (lower panel)} lying southwest of the OT (see Fig. 6) acquired on 
April 7 with VLT-{\it Melipal} plus FORS1. The [O {\sc ii}] $\lambda$3727
and [O {\sc iii}] $\lambda$5007 emission lines, at a redshift $z$ = 0.472, 
are indicated}
\end{figure}

From the detection of [O {\sc ii}] $\lambda$3727 and [O {\sc iii}] 
$\lambda$5007 emission lines in their
spectra, galaxies `1' and `2' appear to be both at redshift 
$z$ = 0.472$\pm$0.001, substantially lower than that of the GRB.
This is the same redshift of one of the absorption systems detected in 
the OT spectrum. Therefore, galaxy `1' is most probably responsible for 
that absorption system. Moreover, galaxy `1' cannot be interacting with 
the GRB host.
This result overrides that reported in Masetti et al. (2002b): the
preliminary analysis of the spectrum of galaxy `1' led to an incorrect 
result because the contamination by the host galaxy of the GRB, at 
$z$ = 0.691, had not been properly removed.

\subsection{Polarimetry}

In order to evaluate the Stokes $Q$ and $U$ parameters of the OT emission,
and thence its polarization percentage $P$ and position angle $\theta$, we
applied the method described by di Serego Alighieri (1997). Here $P$ and
$\theta$ are obtained by using the relation $S(\phi) = P$ cos 2($\theta -
\phi$), in which $S$ depends on the ratio between the fluxes of ordinary
and extraordinary components of the incident beam, and $\phi$ corresponds 
to the prism rotation angle. The values of $P$ and $\theta$ are calculated
by fitting the above relation to the measurements obtained in
correspondence of each rotation angle. Moreover, using this formalism,
one obtains the two Stokes parameters $Q$ and $U$ from the values of
$S(0^{\circ})$ and $S(45^{\circ})$, respectively.
The intrinsic nature of the OT $R$-band polarization is supported by
the relative location of the OT with respect to that of field stars in the 
$U$ vs. $Q$ plot (Fig. 8, upper panel).

After correcting for spurious field polarization, we found
$Q_{\rm OT}$ = 0.016$\pm$0.006 and
$U_{\rm OT}$ = $-$0.009$\pm$0.006. The fit of the data with the relation
described above (see Fig. 8, lower panel) yielded for the OT
a linear polarization $P_{\rm OT}$ = 1.5$\pm$0.4 \% and a polarization
angle $\theta_{\rm OT}$ = 172$^{\circ}$$\pm$8$^{\circ}$, corrected
for the polarization bias (Wardle \& Kronberg 1974).
This latter correction is introduced because $P$ is a positive definite
quantity, and thus at low S/N polarization levels the distribution
function of $P$ is no longer normal but becomes skewed, which causes an
overestimate of the real value of $P$ (Simmons \& Stewart 1985).

\begin{figure}
\psfig{file=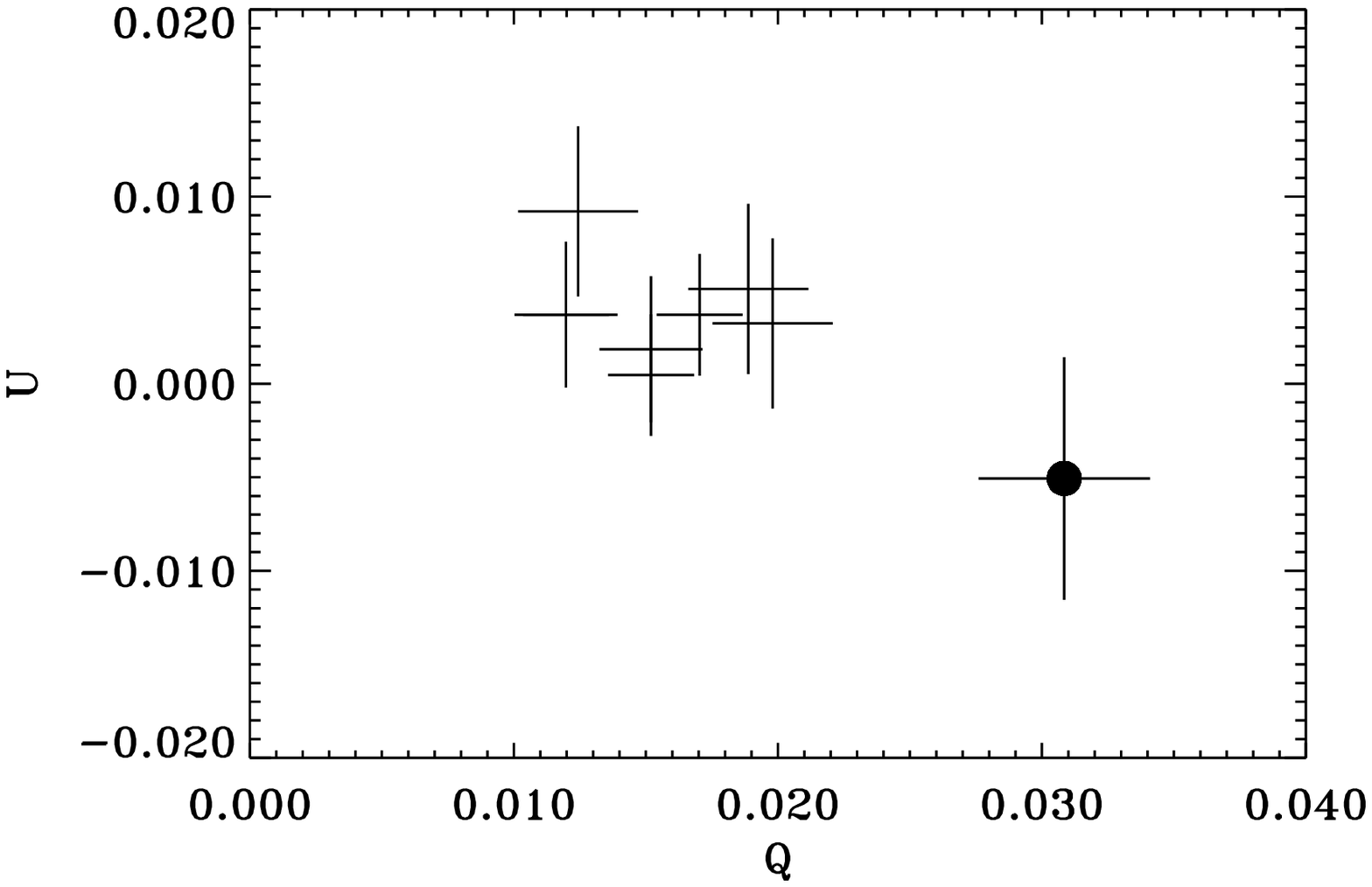,width=8.8cm}
\psfig{file=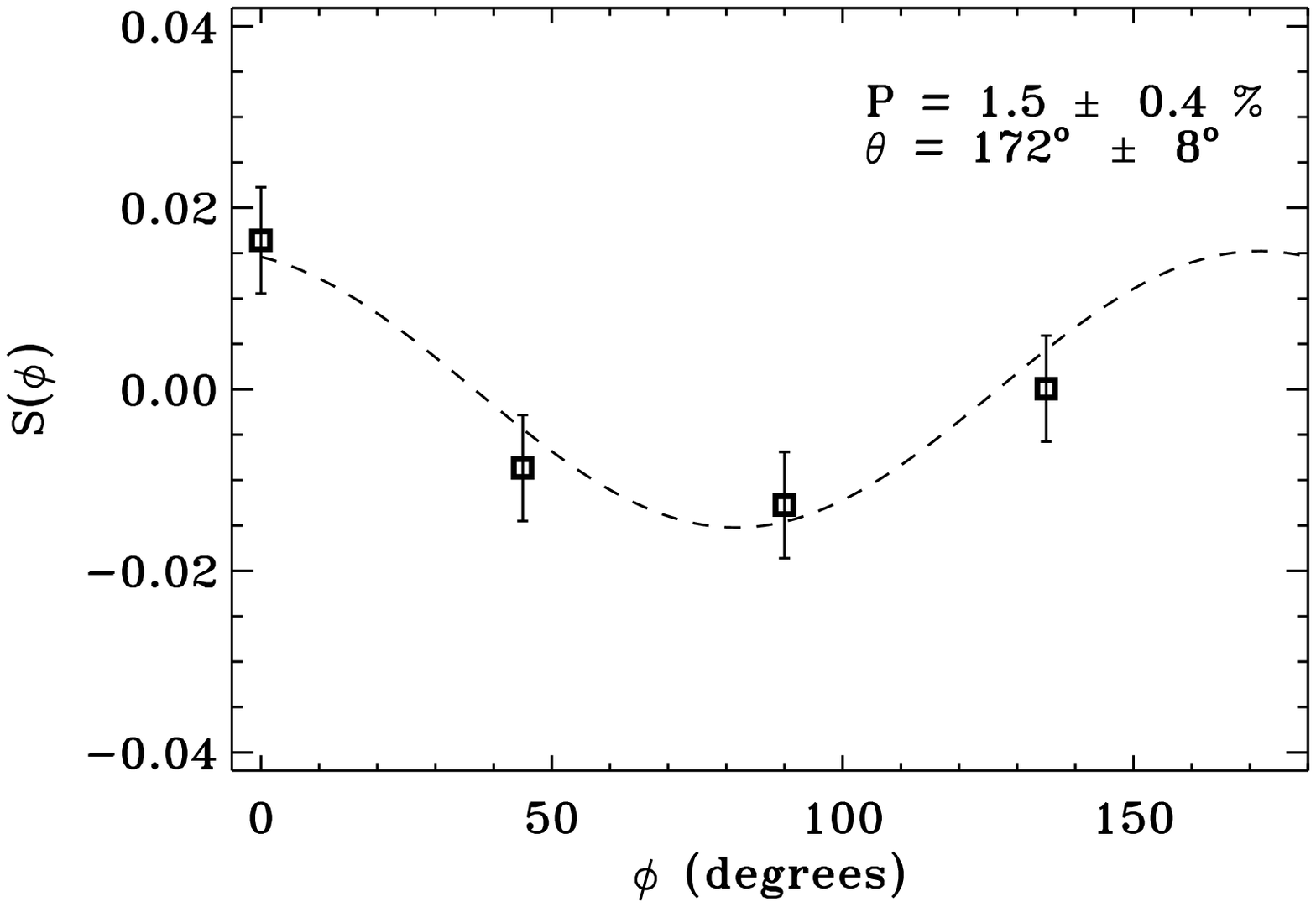,width=8.8cm}
\vspace{-.3cm}
\caption{{\it (Upper panel)}: positions
of field stars and of the OT (marked with a filled dot) in the plane of 
the $U$ and $Q$ parameters not corrected for spurious (instrumental
plus field) polarization. The OT is clearly separated from the region
occupied by the field stars: this indicates that it has net intrinsic
polarization. {\it (Lower panel)}: cosine fit of the $R$-band polarimetric
data of April 6. The best fit yields $P$ = 1.5$\pm$0.4 \% and
$\theta$ = 172$^{\circ}$$\pm$8$^{\circ}$}
\end{figure}

The use of standard polarization equations (e.g., Ramaprakash 1998) 
leads to results which are in good agreement, within the uncertainties, 
with those obtained from the method described above.
In our treatment, we implicitly assumed that the OT polarization degree
and angle did not vary across the four exposures based on the fact that
the total flux varies only by 0.025 mag in this time interval.

The spurious polarization contribution due to the Galactic ISM and
produced by dust grains located along the line of sight (preferentially
aligned in one direction) depends on the $E(B-V)$ color excess according 
to the empirical relation (Hiltner 1956; Serkowski et al. 1975) 
$P_{\rm ISM} (\%) = 9.0 \cdot E(B-V)$. In this case $E(B-V)$ = 0.055, as
mentioned above, so the total ISM-induced polarization is 0.5\%. 
The fraction of this due to the close Galactic ISM has been 
taken into account and corrected by normalizing the OT polarization 
measurement to several field stars, as described in Sect. 2.3. 
Indeed, we find that the overall spurious polarization of the field 
stars is $P_{\rm sp}$ = 0.5$\pm$0.1 \%. This means that it is entirely due 
to the Galactic ISM, which further confirms that the instrumental 
polarization is negligible.
We do not expect that the far (Galactic Halo) ISM contributes
significantly to the OT polarization.
The contribution of the host galaxy to the total (OT+host) light is just
few percent; therefore we do not expect that the host emission 
contaminated significantly our polarization measurement (see also Covino
et al. 2003).
Concerning possible additional local dust absorption within the host (see
the caveats of Covino et al. 2003 about this issue), we will show in
Sect. 4.2 that we do not find evidence for this, based on the analysis of 
the emission of the GRB020405 afterglow. Therefore we tend to conclude 
that the measured OT polarization is intrinsic.

\section{Discussion}

\subsection{Light curves: SN vs. shock re-energization}

The optical light curves of the GRB020405 afterglow are consistent with a
single power-law decay of temporal index $\alpha$ = 1.54$\pm$0.06 between
1 and 10 days after the prompt event, with no temporal break. In the
fireball scenario, this points to a nearly spherically symmetric expansion
(Sari et al. 1998), as also suggested by our fits to the broad-band
spectra of the afterglow (see Sect. 4.2). Our temporal slope is
different than those obtained by Bersier et al. (2003) and Price et al.  
(2003), likely due to the longer time baseline of our sampling and to the
lower host galaxy contribution resulting from our analysis. For the NIR
light curves we find a slightly shallower temporal decay, $\alpha = 1.28
\pm 0.03$.

The red bump detected $\sim$20 days after the GRB has been modeled by
Price et al. (2003) with an emerging SN which is 0.5 mag dimmer than
the Type Ic SN1998bw ($z = 0.0085$) -- considered the ``hypernova" 
prototype (Paczy\'nski 1998, Galama et al. 1998; Patat et al. 2001) --
exploded simultaneously with the GRB at its redshift. Similarly, Dado et
al. (2002) modeled the light curve with a SN akin to 1998bw, but with the
addition of intrinsic extinction. We have also applied a composite
power-law-plus-SN fit to our $VRIJ$ light curves, which extend over a
longer time interval than that considered by the above authors. We first
tried a fit with SN1998bw as a template, redshifted to $z = 0.691$, and
then we used the light curves of another Type Ic SN, 2002ap (Nakano et al.
2002; Gal-Yam et al. 2002b; Pandey et al. 2003), located in M74 (8 Mpc) 
which has also exhibited some hypernova characteristics
(Mazzali et al. 2002; see however Berger et al. 2002). 
SN2002ap showed a faster time evolution after maximum with respect 
to that of SN1998bw, and therefore it could possibly better describe the 
observed post-flattening OT decay.
To construct light curves at the effective wavelengths of the GRB020405
afterglow observations we interpolated the light curves of the two model
SNe or extrapolated to the rest-frame UV range. In order to optimize the
fit, the SN1998bw and SN2002ap temporal profiles were dimmed by 0.6 mag
and brightened by 1.3 mag, respectively.

In Fig. 9 are reported the power-law+SN fits for the case in which the
SN2002ap was used as a template. Fig. 10 shows the fit with a power 
law plus SN1998bw in the $R$ band only. Either using SN1998bw or SN2002ap
as a template, the fit is not satisfactory in the $V$ band. This can be
expected, since this wavelength range corresponds to the near-UV at the
GRB redshift, where SNe suffer severe intrinsic extinction in the
envelope. In the other bands, the fit is marginally satisfactory for both
SN templates.
Allowing for a positive or negative time lag between the SN and GRB
explosions does not produce any significant improvement in the fit.

\begin{figure*}
\hspace{-0.4cm}
\psfig{file=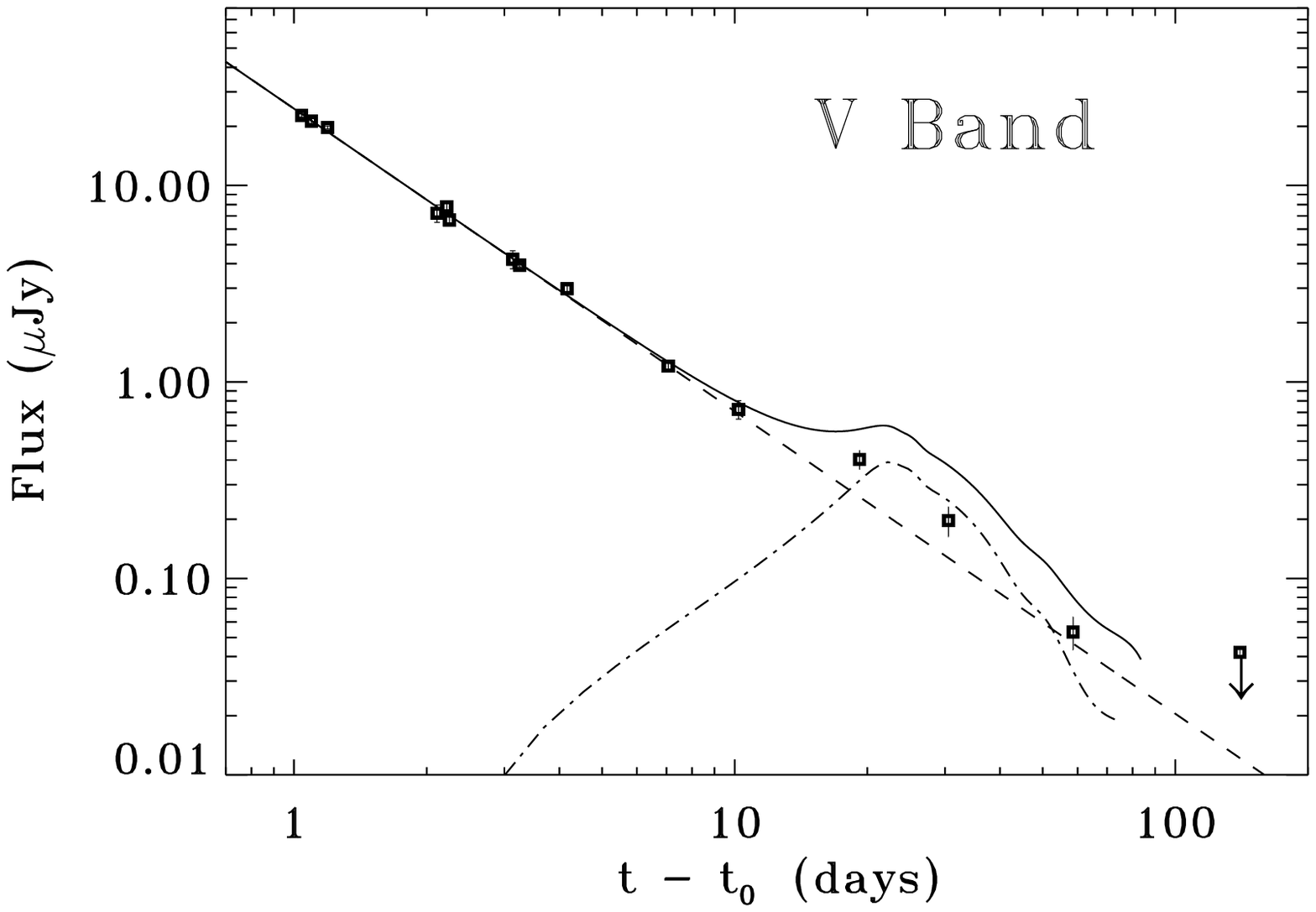,width=9.4cm}
\psfig{file=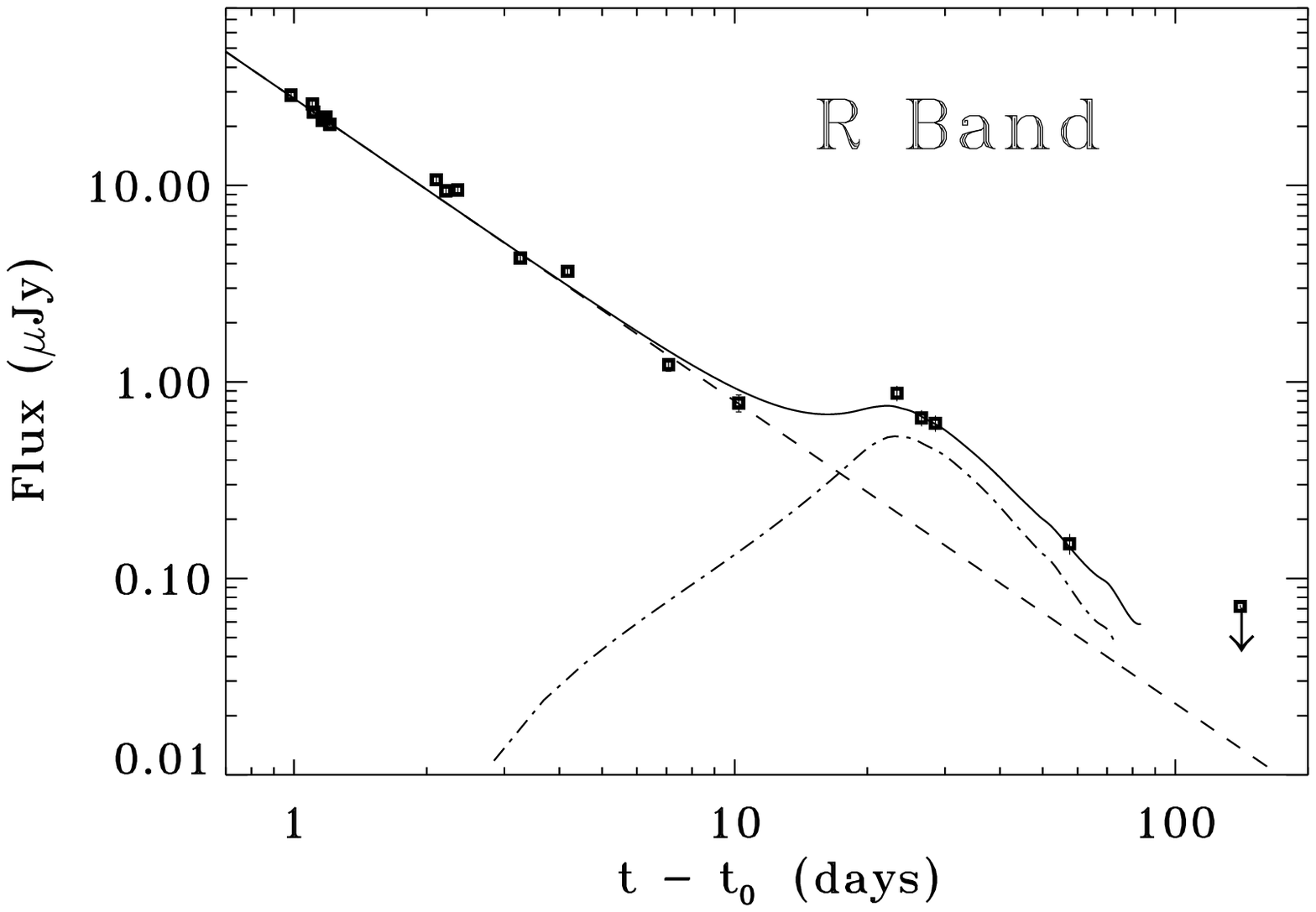,width=9.4cm}

\hspace{-0.4cm}
\psfig{file=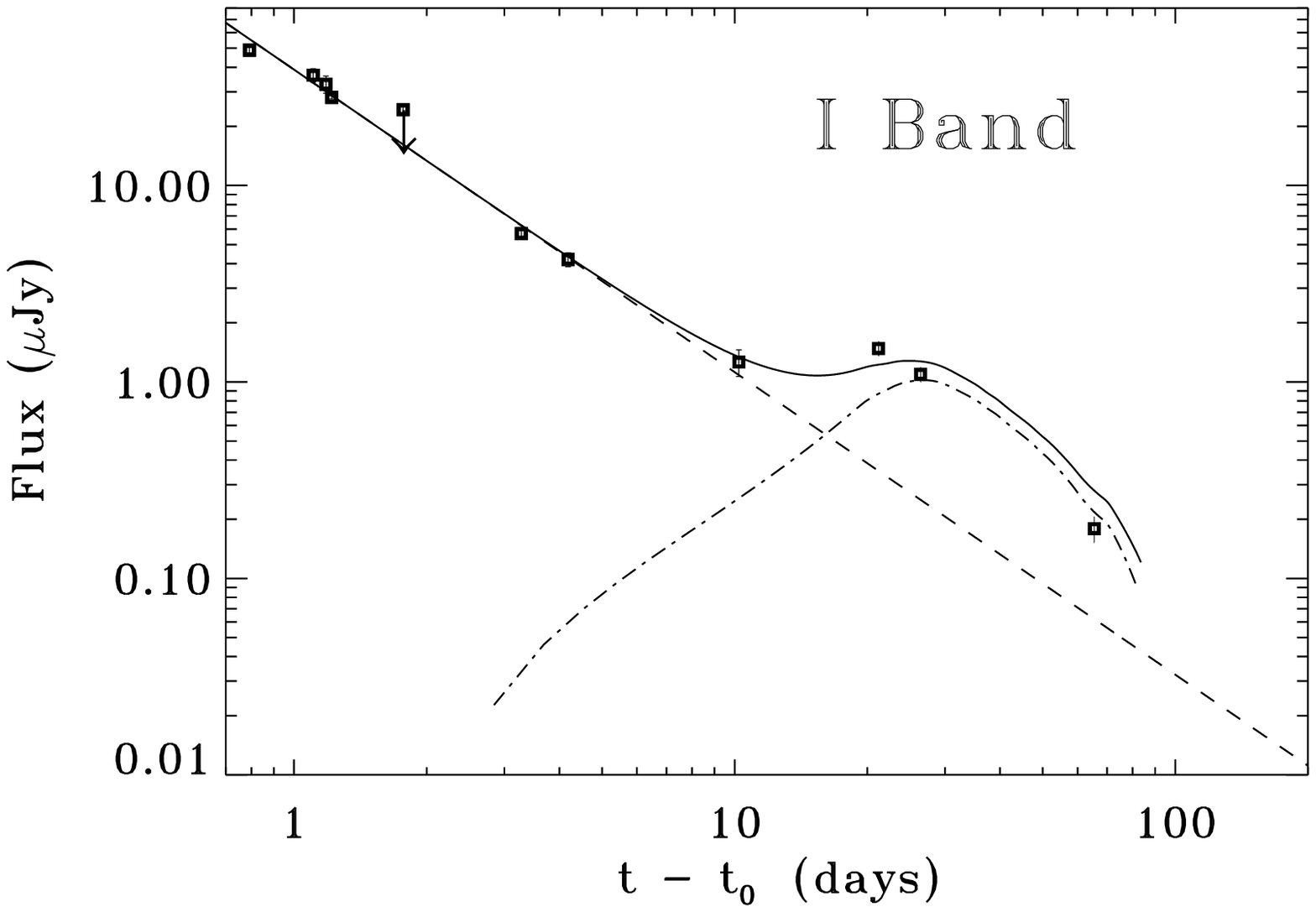,width=9.4cm}
\psfig{file=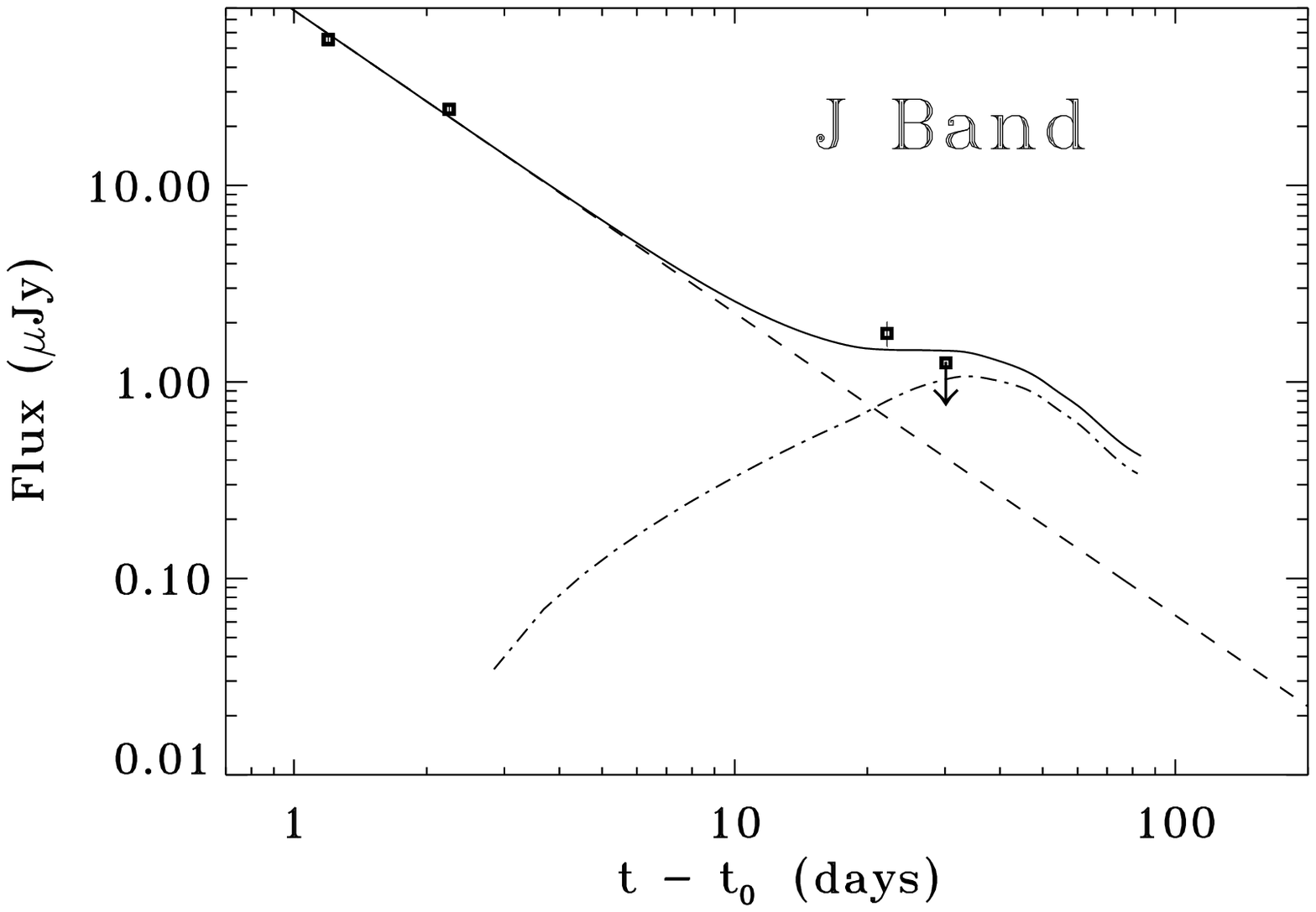,width=9.4cm}
\caption{$V$ (upper left), $R$ (upper right), $I$ (lower left) and  
$J$ (lower right) light curves of the GRB020405 afterglow;
$t_0$ indicates the time of the GRB onset. The data are corrected for the
host galaxy contribution and for Galactic absorption,
and fitted with a power law with $\alpha$ = 1.54 (dashed line) plus a
SN2002ap at $z$ = 0.691 (dot-dashed line) brightened by 1.3 mag with
respect to the original one. The solid line represents the best-fit model.
In these fits the SN was considered as occurred simultaneously with the
GRB. Light curves of the SN2002ap template end at $\sim$80 days after the 
explosion because after that date they are poorly sampled}
\end{figure*}

Alternatively, the afterglow temporal variability may be reproduced by
both density and energy variations (e.g., Nakar et al. 2003).
Specifically, the observed bump may be interpreted within the scenario of
shell collisions proposed by Kumar \& Piran (2000), in which a slower
shell runs into a faster but decelerating shell and re-energizes it.  The
amplitude of the bump depends on the ratio of the energy amounts carried
by the two shells.  The optical light curves of GRB020405 exhibit a larger
temporal slope ($\alpha'$ = 1.85$\pm$0.15) after the bump than
before it ($\alpha$ = 1.54$\pm$0.06), which would imply also a change in
the electron energy distribution shape, an effect which is not predicted
by Kumar \& Piran (2000). While this circumstance may limit the
applicability of their picture to our observations, a similar but more
refined model could be consistent with them. Our light curves
could be also qualitatively reproduced by an analogous scenario proposed
by Ramirez-Ruiz et al. (2001), in which light variations
are caused by the interaction of the blast wave with the density bumps in
a pre-ejected stellar wind. However, this interpretation would not be
self-consistent, since the hydrodynamics of the GRB020405 afterglow points
to a homogeneous, rather than to a windy, circumburst medium (see below). 

Other views are offered by the model developed by Beloborodov (2003), in
which the fireball is interacting with a trailing neutron shell, or by
that in which a SN remnant, located around the GRB progenitor and excited
by the GRB itself, cools down on time scales of weeks (Dermer 2002).
Unfortunately, no thorough quantitative descriptions of the effects
induced by these interactions on the spectral and temporal behaviour of
OTs are available at the moment. 

The dust echo model (Esin \& Blandford 2000; Reichart 2001) seems instead 
to be a less viable interpretation for the emission excess because it 
predicts bluer spectra at the time of the bump compared with those 
displayed by the early-time observations assuming scattered light 
dust echo; this dust echo emission is ruled out by the data.
This model however, in case of thermal dust echo, also foresees a 
possible spectral peak in the observer's frame far-IR, which we 
cannot test. However, the increase in the radio flux noticed by Berger et 
al. (2003) around day 20 after the GRB may be consistent with this 
explanation.
The microlensing interpretation (e.g., Loeb \& Perna 1998) of the
phenomenon is instead unlikely, as any variation is expected to be
achromatic in this case. 

The temporal decay steepening seen in the optical at day $\sim$20 after
the GRB ($\alpha'$ = 1.85) might also be interpreted as a
transition to a non-relativistic regime (Dai \& Lu 1999) for the same
electron distribution shape derived below in Sect. 4.2. However, in 
this case we would not expect significant spectral steepening after the 
transition, as opposed to what is observed.

\subsection{The broad-band spectrum}

Mirabal et al. (2002) propose a wind model for the GRB020405 afterglow
expansion: the blast wave propagates in a medium enriched before the GRB
explosion by a wind ejected from a Wolf-Rayet star. Under the hypothesis
that the cooling frequency is higher than the X--ray energies at the epoch
of their Chandra observations, those authors find, based only on the
X--ray data, an electron distribution slope of $p = 2.7$. However, the
X--ray spectral shape, with a spectral index $\beta_{\rm X}$ = 0.7$\pm$0.2, 
is significantly flatter than the one we derive in the optical,
$\beta_{\rm opt}$ = 1.3$\pm$0.2. If one allows for some intrinsic
extinction within the GRB host galaxy, one may recover the agreement
between the optical and X--ray spectral indices; however, in this case it
would be difficult to explain why the temporal decline rate is different
at the NIR, optical and X--ray frequencies (Chevalier \& Li 2000). If
instead no intrinsic extinction is invoked, the flatness of the X--ray
spectrum makes it impossible to reconcile it with synchrotron emission,
predicted by the standard fireball scenario to be produced as a
consequence of the interaction between the external shock and the
interstellar medium (Sari et al. 1998) or the circumburst medium enriched
by a pre-ejected stellar wind (Chevalier \& Li 2000). Therefore, we favor
an interpretation of the X--rays as Inverse Compton emission, as also
proposed by Mirabal et al. (2002), and as in the case of GRB000926
(Harrison et al. 2001).
Since the X--ray spectral slope is consistent, within the errors,
with the NIR slope, the same population of relativistic electrons may have
produced the NIR photons via synchrotron radiation and the X--rays via
Inverse Compton scattering off afterglow radiation (synchrotron
self-Compton).

The optical and NIR data are inconsistent with a wind afterglow model
(Chevalier \& Li 2000): the spectral break between the NIR and optical
bands (Fig. 4) could not correspond to the synchrotron injection break
($\nu_{\rm m}$), since then the observed NIR slope would be expected to be
much flatter than observed; therefore, this break must be due to the
synchrotron cooling frequency ($\nu_{\rm c}$). However, in this case the
NIR temporal decline should be faster than seen in optical, contrary to
what is observed.
Assuming instead an isotropic and adiabatic expansion in an external
medium of constant density (``homogeneous" model; Sari et al. 1998) and a
slow cooling regime, the spectral slopes bluewards and redwards of the
cooling frequency $\nu_{\rm c} \simeq 2.5 \times 10^{14}$ Hz imply an
electron energy distribution index $p$ = 2.5$\pm$0.3. This is consistent
with the (better constrained) value, $p$ = 2.7$\pm$0.1, obtained from the
NIR and optical decay rates under the assumptions of the homogeneous
model with no need for additional local absorption in the host.

From this, and from the above considerations on the broad-band 
X--ray-optical-NIR spectral flux distribution, we conclude that the 
fireball emission is affected by negligible wavelength-dependent 
extinction associated with the host galaxy.
Similarly, the OT detection in the $U$ band ($\lambda_{\rm restframe} 
\sim$ 2100 \AA) suggests, in a model-independent way, negligible intrinsic 
absorption.

We note that a collimated fireball before the jet break (Sari et al.
1999; Rhoads 1999) and an isotropic fireball (Sari et al. 1998) are
described in equivalent ways by the standard model and that they cannot
be discerned by the temporal and spectral indices. If the bump observed
in the light curves at day $\sim$20 is due to an emission component
independent from the afterglow (i.e., a SN), our data cannot rule out a
steepening of the afterglow light curve, related to a collimation
break, occurring more than $\sim$10 days after the GRB. According to
Sari et al. (1999), this would imply a lower limit to the jet opening
angle of $\Theta > 14^\circ$.

The conclusion that the GRB020405 fireball expanded in a homogeneous
medium was independently reached by Berger et al. (2003) from
observations of the radio afterglow. However, these authors also suggest
the presence of a jet break occurred $\approx$1 day after the GRB.
While our optical-NIR light curve coverage does not allow us to confirm
their finding, this interpretation would imply that we monitored the
counterpart decay during the post-break phase of a collimated fireball.
Following Sari et al. (1999), this would require that $p \equiv \alpha
\sim$ 1.5, which is inconsistent with our spectral results.

\begin{figure}
\hspace{-0.4cm}
\psfig{file=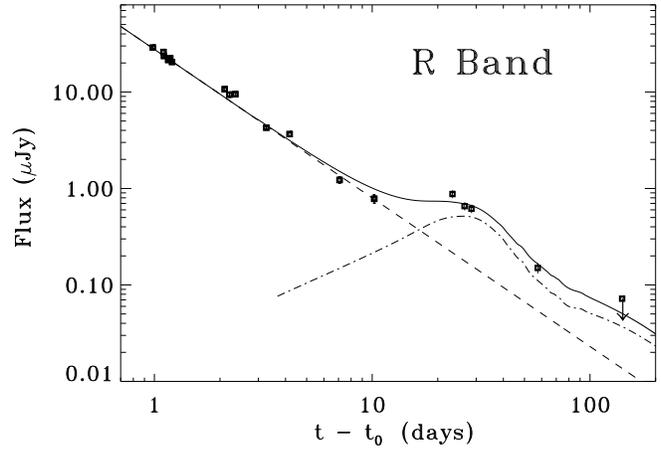,width=9.4cm}
\caption{Same as the upper right panel of Fig. 9, but with SN1998bw as a
template, dimmed by 0.6 mag}
\end{figure}

\subsection{Optical polarization}

Our $R$-band polarization measurement is in line with the detections
and upper limits so far obtained for GRB optical afterglows (Hjorth
et al. 1999; Wijers et al. 1999; Covino et al. 1999; Rol et al. 2000;
Bj\"ornsson et al. 2002; Covino et al. 2002c; Covino et al. 2002d; 
Covino et al. 2002e; Rol et al. 2003).

Jet geometries are thought to produce observable polarization (Sari 1999; 
Ghisellini \& Lazzati 1999), however, no indication of a jet structure is
derived from the light curve analysis for the GRB020405 afterglow (see
Sects. 3.1 and 4.1). These characteristics, i.e., the presence of
detectable polarization and the absence of a light curve steepening
(at least in the first 10 days), are reminiscent of GRB990712 
optical afterglow (Rol et al. 2000; Sahu et al. 2000), although a jet 
break has been suggested by Bj\"ornsson et al. (2001) for the latter. 

Our values of $P$ and $\theta$ are consistent with those measured for
the present afterglow by Covino et al. (2003) in the $V$ band
$\sim$2 and $\sim$3 days after the GRB. However, our polarization
percentage measurement is at variance with that obtained by Bersier
et al. (2003) from $V$-band observations acquired nearly
simultaneously with ours. This cannot be explained by the contribution
from the underlying host galaxy, because (as we remarked in Sect. 3)
this is negligible at the time of our polarimetric
observation. Also, a strong dependence of the polarization value on
the wavelength is unlikely (Rybicki \& Lightman 1979) and, indeed,
recent optical spectropolarimetric observations of the afterglows 
of GRB020813 (Barth et al. 2003) and GRB021004 (Wang et al. 2003) show no
substantial variations of $P$ across the continuum redwards of 4000 \AA.

For a spherically symmetric fireball, the net polarization should vanish
if the magnetic field is ordered exactly perpendicular or
parallel with respect to the expanding shock front (e.g., Waxman 1997). 
Therefore, rapidly variable effects removing the symmetry should be
considered in order to explain the different polarization values measured
by us and by Bersier et al. (2003), like turbulence or microlensing
(Medvedev \& Loeb 1999). In these cases substantial variations of
$\theta$ would also be expected (see Rol et al. 2000 and references
therein), but they are not observed. Therefore, if real, this fast
variation of $P$ lacks a satisfactory interpretation.

Thus, as also pointed out by Rol et al. (2000) for GRB990712, no current
polarization model adequately explains the entire optical polarimetry data
set for the GRB020405 afterglow, that is, the strong variability of $P$
and the simultaneous constancy of $\theta$ when no break is seen in the
optical light curve during the polarimetric monitoring (see however 
Bj\"ornsson \& Lindfors 2000).
We independently analyzed the polarimetric data of Bersier et al.      
(2003) and, by applying our PSF-fitting photometric procedure, which
differs from that based on the ISIS2 package (Alard 2000) used by Bersier
et al. (2003), we obtain a 3-$\sigma$ upper limit on the $V$-band
polarization of 36\%. This value is higher than, but still consistent
with the measurement and upper limit obtained by the above authors.

\subsection{The host galaxy and the intervening absorption}

The ground-based and HST imaging reveal that the GRB020405 OT is
located in a complex field (see Figs. 3 and 6). Emission and absorption
lines detected on the optical spectra of the OT indicate a redshift
of $z$ = 0.691, confirming our preliminary analysis (Masetti et
al. 2002a) and the independent spectroscopic results by Price et
al. (2003). An absorption system at the lower redshift of $z$ = 0.472
is also detected in the OT spectrum acquired on April 7 (see
Fig. 5), and we attribute it to galaxy `1' (Fig. 6), for
which we detect [O {\sc ii}] and [O {\sc iii}] in emission at the
same redshift in the VLT spectrum.  At this redshift the
angular separation ($\sim$2$''$)  between galaxy `1' and the OT
corresponds to $\sim$13 proper kpc, consistent with the
halo of galaxy `1' being responsible for the absorption system at $z$
= 0.472 detected in the OT spectrum. For galaxy `2', located at
$\sim$6$''$ from the OT (Fig. 6), we also measured a redshift $z$ = 0.472,
therefore the two galaxies probably belong to a foreground association or
cluster. Galaxy `2' may also be responsible for the absorption system,
although its larger angular distance from the OT makes it a less probable
candidate than galaxy `1' for the absorber. The presence of
intervening absorption systems is not uncommon in the OT spectra (e.g.,
Metzger et al. 1997; Masetti et al. 2001; Barth et al. 2003; M\o ller et
al. 2002). However, the present case is the first for which the
intervening absorber has been identified with imaging and spectroscopy.

The flux of the [O {\sc ii}] $\lambda$3727 can be used to determine the
star formation rate (SFR) of the GRB020405 host galaxy. Applying Eq. 3 of
Kennicutt (1998), we determine a SFR of $\sim$3 M$_\odot$ yr$^{-1}$,
consistent with the findings of Price et al. (2003). As noted by these
authors, this is actually a lower limit to the SFR as no correction for
the unknown local absorption has been determined.

The detection of [O {\sc ii}], [O {\sc iii}] and $H_\beta$ emission lines
in the spectrum also allowed us to infer the oxygen abundance in the
GRB host. By computing the parameter $R_{\rm 23}$, defined by Kobulnicky
et al. (1999) as the ratio between the sum of the oxygen forbidden line
fluxes and the $H_\beta$ line flux, we find $R_{\rm 23}$ = 6.5$\pm$2 which
implies, from Fig. 8 of Kobulnicky et al. (1999), that 12 + log (O/H) =
8.0$\pm$0.25. This suggests that the GRB020405 host has an
oxygen abundance lower (around one eighth) than solar; also, this
metallicity yield is similar to those found for other GRB hosts (e.g., 
Price et al. 2002c).

\section{Conclusions}

The most important results of our ground-based observing campaign on the
GRB020405 afterglow can be summarized as follows: 

\begin{itemize}

\item
we have reported the first detection of the NIR afterglow of this GRB: the
OT is clearly detected in $JHK_s$ bands;

\item
thanks to our accurate method of host subtraction we find that, between 1
and 10 days after the GRB, the decay of the afterglow light curves is
consistent with a single power law. Therefore we model the afterglow
emission as synchrotron radiation in a fireball expanding in a homogeneous
medium and within a jet whose opening angle is not smaller than
14$^\circ$. At X-ray frequencies, the 
contribution of synchrotron
self-Compton radiation may dominate; 

\item
the addition of late-time (20-150 days after the GRB) HST points and VLT
$J$-band observations indicates the presence of a red bump in the
$VRIJ$ light curves around 20 days after the GRB. This bump can be
modeled with a SN. Alternatively, HST and late-time $J$-band data can be
modeled by using a power law with decay index steeper than that of the
early decline phase ($\alpha'$ = 1.85$\pm$0.15). This can be explained as
due to a late shell collision in the fireball; 

\item
a $R$-band polarimetry measure shows that the afterglow is polarized, with
$P$ = 1.5$\pm$0.4 \% and $\theta$ = 172$^{\circ}$$\pm$8$^{\circ}$. No
quantitative model can explain the entire polarization data set for this
OT; 

\item
the spectroscopic detection of emission lines indicates that the GRB is
located at redshift $z$ = 0.691. Two metallic absorption systems, at $z$ =
0.691 and at $z$ = 0.472, are also detected in the afterglow optical
spectrum. This latter redshift coincides with that of a galaxy complex
angularly close (2$''$) to the host of GRB020405, indicating that the
absorption system most likely originated in this complex. For the first
time, the galaxy responsible for an intervening absorption line system in
the spectrum of an OT is directly detected; 

\item
the host of GRB020405 appears to be a galaxy with moderate SFR ($\sim$3
M$_\odot$ yr$^{-1}$) and with subsolar metallicity;

\item
while monitoring the late phases of the afterglow decay, HST revealed the
presence of a further variable object located $\sim$3$''$ from the OT, in
the southern outskirts of galaxy `1'. While its nature is still to be
investigated, its presence underscores the need for a careful validation
of the variable candidates detected in GRB error boxes as counterparts of
the high energy events.

\end{itemize}

Our results point to the need of further systematic NIR/optical campaigns
of GRB afterglows, coordinated with higher energy monitorings, to
better constrain the emission mechanisms and the environmental
characteristics of GRBs and of their lower energy counterparts.

\begin{acknowledgements}

We are grateful to the referee, Dr. E. Le Floc'h, for several useful 
comments which helped us to improve the paper.
We thank R. Silvotti, D. Stephens and D. Thomas for having kindly allowed
us to acquire some of the data presented here during their observational
time. We also thank the ESO staff astronomers M. Billeres (La Silla) and
H. B\"ohnhardt, R. Cabanac, A. Kaufer, R. Johnson and A. Jaunsen (Paranal) 
as well as J. Licandro (TNG), T. Augusteijn and R. Greimel (WHT and JKT)
for their help and efforts in obtaining the observations presented in this
paper. The TNG data presented here were acquired in the framework of the
Italian GRB Collaboration at TNG. Paul Price is thanked for having swiftly
provided us the position of the OT soon after its discovery, as well as
the June 2002 HST data prior to their public release. 
We acknowledge Scott Barthelmy for maintaining the GRB Coordinates Network
(GCN) and BACODINE services. This research has made use of NASA's
Astrophysics Data System and of the Multimission Archive at the Space
Telescope Science Institute (MAST). This work was supported by the Danish
Natural Science Research Council (SNF).
J.M. Castro Cer\'on acknowledges the receipt of a FPI doctoral fellowship 
from Spain's Ministerio de Ciencia y Tecnolog\'{\i}a.
\end{acknowledgements}

\end{document}